\documentstyle[11pt,aaspp4]{article} 
\newcommand{\kms}{km~s$^{-1}$}
\newcommand{\teff} {$T_{\rm eff}$\/}

\newcommand{\etal}{{\it et al.\/}}
\newcommand{\eqw}{$W_{\lambda}$}

\newcommand{\ep}{$\chi$}

\newcommand{\mc}[1]{\multicolumn{2}{c}{#1}}

\begin{document}

\title{An Abundance Analysis for Four Red Horizontal Branch Stars in the
Extremely Metal Rich Globular Cluster NGC 6528\altaffilmark{1}}

\author{Eugenio Carretta \altaffilmark{2},
        Judith G. Cohen\altaffilmark{3},
        Raffaele G. Gratton\altaffilmark{2} \& 
        Bradford B. Behr\altaffilmark{3,4}}

\altaffiltext{1}{Based in large part on observations obtained at the
	W.M. Keck Observatory, which is operated jointly by the California 
	Institute of Technology and the University of California}
\altaffiltext{2}{Osservatorio Astronomico di Padova, Vicolo dell'Osservatorio 5,
        35122, Padova, Italy}
\altaffiltext{3}{Palomar Observatory, Mail Stop 105-24,
	California Institute of Technology, Pasadena, CA \, 91125}

\altaffiltext{4}{Current address: University of Texas, Department of Astronomy,
Austin, Texas 78712}

\begin{abstract}

We present the results of the first analysis of high dispersion spectra of four
red HB stars in the metal rich globular cluster NGC 6528, located in Baade's
Window. We find that the mean [Fe/H] for NGC 6528 is $+0.07 \pm 0.01$\ dex
(error of the mean),
with a star-to-star scatter of $\sigma = 0.02$\ dex (4 stars), although the 
total error is likely to be larger ($\sim 0.1$ dex) due to systematic
errors related to the effective temperature scale and to model atmospheres.
This metallicity is
somewhat larger than both the mean abundance in the galactic bulge found by
McWilliam \& Rich (1994) and that found in our previous paper for NGC 6553.
However, we find that the spectra of clump stars in NGC 6528 and NGC 6553 are
very similar each other, the slightly different metal abundances found being
possibly due to the different atmospheric parameters adopted in 
the two analyses.
Since the present analysis is based on higher quality material, we propose
to revise our previous published metal abundance for NGC 6553 to 
[Fe/H]=$-0.06 \pm 0.15$.

For NGC 6528 we find excesses for the $\alpha$-process elements Si and Ca
([Si/Fe]=+0.4 and [Ca/Fe]=+0.2), whereas Mn is found to be 
underabundant ([Mn/Fe]=$-$0.4). We find a solar abundance of O; however this
is somewhat uncertain due to the dependence of the O abundance on the adopted 
atmospheric parameters and to coupling between C and O abundances in these
cool, metal-rich stars. Finally, we find large Na excesses ([Na/Fe]$\sim
+0.4$) in all stars examined.

\end{abstract}


\section{Introduction}

The bulge is one of the major components of the Milky Way. Its integrated
properties resemble those of giant ellipticals, which in turn are one of the
dominant components of the Universe. However, in contrast to these distant
environments, the relative proximity of the galactic bulge allows us to
study individual stars in detail, and (at least in some cases)
derive accurate abundances for the local version of old metal rich
populations. This capability is very helpful in reconstructing the
history of formation, still quite controversial, of these very important
constituents of the Universe.

Unfortunately, owing to the rather large distance and to substantial
interstellar absorption characteristic of the bulge, without a significant
boost from microlensing, detailed abundance analyses of bulge stars have to be
limited to the evolved population. An exploratory study of field bulge stars
was carried out by McWilliam \& Rich (1994); coupled with other studies at
lower resolution (Rich 1988), this study indicated that the bulge has a wide
abundance distribution, peaking at a metal abundance (as represented by
[Fe/H]) slightly lower than solar, and may have an excess of
$\alpha-$elements. The latter would suggest that most of the bulge formed
within a rather short interval of time, due to the interplay between star
formation and the lifetime of the progenitors responsible for the synthesis of
different elements.

While this analysis is of great value, the interpretation of the results is
complicated by the fact that we are not able to determine the ages of the
stars observed.  Thus, for example, from these data alone, we are unable to
deduce any strong constraint on the epoch of formation of the bulge. This
concern may at least in principle be overcome by studying stars in clusters.
The bulge of our Galaxy has a rich population of globular clusters (Ortolani
1999). The colour-magnitude diagrams (CMDs) of these clusters indicate that
several of them are metal-rich, (quite similar to the bulk of the field bulge
population), and they are likely old (Ortolani \etal\ 1995). However, accurate
metal abundances are needed to determine ages with the precision required to
understand if the bulge is as old as (or even older than) the halo, or if
instead it has  the somewhat younger age of the oldest stars in the thin disk.
Thus, accurate abundance determinations for bulge globular clusters represent
a basic step in our understanding the formation of the Milky Way.

Such an analysis also provides additional important pieces of information,
since it allows us to: (i)  extend the calibration of abundance scales at high
metal abundance, poorly known at present since metal-rich clusters tend to be
concentrated toward the galactic center, where reddening and crowding often
hamper accurate observations; (ii) derive quite accurate reddening estimates by
comparing the observed colours with those expected for stars having
temperatures determined from spectroscopically derived parameters, such as line excitation, which are independent of reddening: and (iii) further constrain the
evolution of the bulge by determining the ratios of the abundances of
$\alpha-$elements to iron (crucial to determining the rate of chemical
evolution).

In order to address these questions, in a previous study (Cohen \etal\ 1999,
henceforth Paper 1) we used the Keck Telescope to acquire high resolution
spectra of individual stars in NGC~6553. The large aperture of the Keck
telescope and the efficiency of its high resolution spectrograph (HIRES, Vogt
\etal\ 1994) allowed us to observe stars on the red horizontal branch (RHB). 
As explained in Paper I, this choice is clearly advantageous as compared to
observations of first ascent red giants, since RHB stars are warmer, making an
abundance analysis of their spectra much easier.  Furthermore, contamination
of the RHB region of the CMD diagram by field stars is much less important
than in the case of red giants, resulting in a higher probability of 
membership (as confirmed a posteriori by the measured radial velocities). In
Paper I we found that NGC~6553 has a metallicity of [Fe/H]=$-0.16\pm 0.08$,
somewhat higher than determined by analysis of two cool giants by Barbuy
\etal\ (1999, henceforth B99), and very similar to the bulk of bulge field
giants observed by McWilliam \& Rich (1994). Also, we found an excess of the
$\alpha-$elements similar to that found for the field stars in the bulge.

In the present paper we perform for the first time a similar study on
NGC~6528, a globular cluster having values of the metallicity indicators very
similar to those of NGC~6553 (Harris 1996). NGC~6528 is a very highly
concentrated cluster, between us and the galactic bulge, at about 7.8 Kpc from
the Sun, hence very close to the galactic center (e.g. Ortolani \etal\ 1995).
Located in Baade's Window with $(l,b) = (1.1, -4.2$ deg), it seems
certain that NGC 6528 is  a bulge cluster. Moreover, the cluster 
velocity indicates that it is not a disk cluster.
Ortolani \etal\ (1995) found from
careful comparison of both CMDs and luminosity functions that its stellar
population is very similar to that of NGC 6553 and of Baade's Window.
In addition, recently Davidge (1999), noting that stars in NGC 6528 populate the
same region in the two-colour plane (J-H, H-K) of the field bulge giants,
supports the classification of this cluster as a true bulge cluster, rather
than a very metal-rich disk cluster.

The metallicity of NGC 6528 is uncertain. Zinn \& West (1984) 
derived an abundance of  +0.2 dex, but Armandroff \& Zinn (1988), using an
analysis of the Ca IR triplet in the integrated cluster spectrum, decreased
that value significantly, to $-0.23$ dex.  Other methods, as that of
Sarajedini (1994), cannot be applied due to the lack of calibrating clusters
in the high metallicity regime.

NGC~6553 and NGC~6528 are the only very high metallicity globular clusters that
can be studied rather easily at optical wavelengths; the remaining bulge
clusters are either more metal-poor or very obscured. Hence the choice of
NGC~6528 for the present study was obvious. While the analysis presented here
is very similar to that already done on NGC~6553, it is based on higher S/N 
spectra; moreover, we adjusted the instrumental
set up in order to include lines of Na as well as O. 
This was deemed important because observations of O and Na in very metal-rich
clusters may help to better understand the pattern of abundances of these
elements found in more metal-poor globular clusters (Ivans \etal\ 1999; Kraft
\etal\ 1998). In fact, serious doubts were cast on the up-to-now most favoured 
mechanism (deep mixing) by Gratton \etal\ (2001), who found that the O-Na
anticorrelation (previously seen only for stars on the red giant branch of
globular cluster) is clearly present among dwarf, unevolved stars in NGC 6752.
On the other hand, the alternative explanation suggested (pollution due to mass
lost by previous AGB stars in the cluster) is expected to be sensitive to the
overall metal abundance of the cluster (Ventura \etal\ 2001).
Determination of O abundances in NGC 6528 RHB stars is favoured by the
large radial velocity of this cluster, which shifts stellar lines away from
telluric lines.

\section {Observations and reduction }

\subsection{Choice of Program Stars}

Program stars were selected using high resolution images and VIJK photometry
of NGC~6528 from HST (VI) and IRAC2 (JK) kindly provided by Montegriffo (1999,
private communication).  Table 1 gives the most relevant parameters of these
stars, while Figure~1 shows their location within the cluster and Figure~2 
the position of these objects in the $V,V-I$
color-magnitude diagram of NGC 6528. 
In Figure~1, the stars studied here are marked on this subset from a 100 sec
WFPC2 image from the HST Archive\footnote{Obtained from the data archive at the
Space Telescope Science Institute. STScI is operated by the Association of
Universities for Research in Astronomy, Inc. under NASA contract NAS 5-26555}.
Coordinates for the program stars are given in Table 1.
The field of the cluster is very crowded.
We selected for observation a number of apparently uncrowded stars having
colours and magnitudes appropriate for the RHB in the CMD, in order to
maximize the probability of cluster membership. Beside the 4 program stars
(namely, 5422, 3014, 3025 and 3046) two other stars fell in the edge
of the slit (stars 3032 and 5425) on at least some of the spectra. While not
good enough for abundance analysis, spectra of these extra stars are adequate
for precise radial velocities. {\it A posteriori}, membership of the
program stars in NGC~6528 was confirmed by their radial velocity: this is a
very useful criterion in this case, due to the large $v_r$ of this cluster;
according to Harris (1996), the heliocentric $v_r$ is $184.9 \pm 3.8$ 
(internal error) \kms, and its $v_r$ relative to the Solar local standard of 
rest is 195 \kms. 
Heliocentric $v_r$ for the program stars as measured on our
spectra are given in the column 6 of Table 1: the average heliocentric $v_r$
from our spectra is $209.9 \pm 1.6$ \kms\, with 
$\sigma = 4.0$ \kms\ from 6 stars, giving double weight to velocity from the
June 2000 spectra that have a higher S/N ratio.

The radial velocities were measured by cross correlating the region
6130 - 6170 \AA\ in echelle order 57 and the region 6240 - 6280 \AA\
in echelle order 58, using the June 2000 spectrum of star 3025 as
a template. 
These measurements were carried out independently and
with different software packages by EC and by JC.
The zero point was determined by fitting Gaussians to
16 lines in these orders in the spectrum of the template
judged to be unblended based on their FWHM.  The laboratory
wavelengths of these lines were taken from the NIST Atomic Spectra Database.
(NIST Standard Reference Database No. 78).  The agreement between
orders is excellent and confirms the dispersion solution from the
Th-A lamp.

The internal radial velocity errors were calculated following
the method of Tonry \& Davis (1979) using the relation
$\sigma(v_r) = \alpha/[1+R(TD)]$, where the parameter $R(TD)$
is a measure of the ratio of the
height of the peak of the cross correlation to the noise in the
cross correlation function away from the peak.
The constant $\alpha$ was set at 15 \kms, which represents a value
typical of those found in other recent HIRES programs using similar
instrumental configurations by
Mateo \etal\ (1998), Cook \etal\ (1999), and C\^ot\'e \etal\ (1999).
The maximum error in $v_r$ incurred by a point source which is
on the edge of the 1 arcsec slit compared to an object that
uniformly fills the slit is 4 \kms, and this factor will
be smaller for real stars under real seeing conditions that are
partially in the slit.  The systematic errors are thus generously
set at 1.5 \kms\ for the program stars observed in 2000,
and 2.5 \kms\ for the stars observed in 1999 as well as for those
that fortuitously appeared in the slits.

Our velocity dispersion measured for NGC 6528 is by far the smallest published
for this cluster, and is consistent with a normal mass-to-light ratio for 
NGC 6528.
However, our mean $v_r$  is
distinctly higher that the literature value in the compilation by Harris.
A search of the references quoted by Harris revealed that all low
values of $v_r$ for NGC 6528 are from older studies, while the most recent
works tend to give a value quite similar to our own. For example, Minniti
(1995) found an average $v_r$ of $203 \pm 20$ \kms\ from 7 stars, noting also
that $v_r$ of this cluster was found to have some discrepancy in previous
analyses (e.g. in Armandroff \& Zinn 1988). The recent extensive work by
Rutledge \etal\ (1997a) found $v_r$ = $212.2$ \kms\ with $\sigma = 13.5$ \kms\
(external error) based on 8 stars. Both these results are in very good
agreement with our $v_r$.  We note that for NGC~6553, 
Rutledge \etal\ (1997a) found $v_r$ =
8.4 \kms\ ($\sigma =8.4$ \kms), in quite good agreement with the value of
Paper I. Moreover, Rutledge et al. (1997a) already noted that their high value
for NGC 6528 was different at the 3.4$\sigma$ level from the mean value in
Harris' compilation and expressed concern about the potential
impact on measurements of $v_r$ of
possible non-members in clusters so heavily contaminated by field stars. This
discussion further supports our adopted strategy of selecting target objects
among the RHB stars.

The confirmation of this high $v_r$ is important as it is among the largest in
absolute value for a bulge cluster so close to the galactic center. 
The radial velocity of NGC~6528 ($209.9\pm 1.6$~km/s) might be a reflection
of the velocity dispersion of the metal rich bulge globular cluster
population (C\^ot\'e 1999) although it appears as rather extreme
for a bulge object: in fact this value is slightly less than 2~$\sigma$\ (of
the scatter of values for individual field stars) from the mean value for
bulge K-giants (Terndrup \etal\ 1995), M-giants (Sharples \etal\ 1990), RR
Lyrae's (Gratton \etal\ 1987), and Miras (Feast \etal\ 1980). While this large
radial velocity clearly rules out the possibility that NGC6528 is a disk
cluster, there is the possibility that it belongs to the inner halo rather
than to the bulge and its orbit is by chance just passing through the bulge at
the present time.
The distinction between inner halo and bulge is not clearly
defined at present. To obtain better insight into this issue, we examined more
closely the large sample of bulge K-giants located in Baade's Window, as is
NGC~6528, observed by Terndrup \etal.
We will consider only those stars fainter
than $V=16$\ (contamination by disk interlopers being important for brighter
objects). For these stars, Terndrup \etal\ found an average velocity of $-8\pm
6$~km/s, with a dispersion of $110\pm 10$~for individual objects. 20 out 334
stars (that is, 6\%) have radial velocities in excess of 200 km/s (in absolute
value), i.e. have kinematics similar to or more extreme than that of NGC~6528.
The average metallicity for these stars (Sadler \etal\ 1996) is
[Fe/H]=$-0.17\pm 0.12$, with a scatter for individual stars of 0.53~dex. The
average value is not significantly different from the average metal abundance
they found for the whole sample of bulge K-giants ([Fe/H]=$-0.11\pm 0.04$).
These stars are clearly much more metal-rich than the traditional value for
halo stars ([Fe/H]$<-1$), and have a metal abundance quite similar to that we
found for NGC6528.
Note that these considerations are based on the high radial velocity of NGC
6528, which makes its membership in the disk population unlikely. On the other
hand, the low radial velocity of NGC 6553 is compatible with that of a disk
object. Some support for this view is given by a recent study by Beaulieu \etal\
(2001). Their color-magnitude diagram for NGC 6553 does not show a good match 
to the field bulge population and favors a metallicity comparable to solar, in 
good agreement with the revised value that we suggest in the present work 
(see below).

\subsection{Details of the Observations}

Observations were carried out with the HIRES spectrograph at Keck I. The HIRES
detector is not large enough to yield full spectral coverage in a single
exposure. The RHB stars in NGC~6528 are faint, and hence the exposures are
long, so a single compromise instrumental configuration is mandatory.  In our
first effort described in Paper I, we wished to avoid crowding of lines, and
hence centered the spectra rather far toward the red.  It turned out that,
ignoring Fe I absorption lines, there were few useful features beyond 8000 \AA\
and also that line crowding was tolerable even at the blue end of the HIRES
spectra of the RHB stars in NGC 6553.  Thus for the observations of the NGC
6528 RHB stars, the instrumental configuration was set to shift the spectra
blueward, so that the O triplet at 7770 \AA\ appears in the reddest order
included. This configuration had the advantage of adding features of several
important elements with no lines in the wavelength regime covered in Paper I,
such as Na, as well as important additional Fe II lines to improve the
analysis of ionization equilibrium. However, the downside of this layout was
that the unblended Mg I line at 8717 \AA\ was lost, while the Mg I lines
included in this setup are more blended and more saturated.

These fields are rather crowded.  Given the freedom to rotate the slit to a
desired position angle and the ability to track at a fixed position angle, in
many cases it might be possible to get two or more candidate RHB stars
within the length of
the HIRES slit. At the time of our initial observations for NGC 6553, HIRES
did not have an image rotator to compensate for the rotation of the field at
the Nasmyth focus, and hence only one star could be observed at a time.
However, by 1999, the HIRES image rotator had been completed by David Tytler
and the Lick Observatory engineering group (Tytler 2000, private
communication). The maximum slit length that can be used with our instrumental
configuration without overlapping echelle orders is 14 arcsec.  Leaving room
at the ends of the slit for sky, we therefore searched the list of RHB
candidates for suitable pairs of RHB candidates located not more than 7 arcsec
apart.

The observations were carried out in two runs 10 months apart:

a) run of August 1999. Due to a hardware problem, the lower dome shutter of
the Keck I telescope was not movable, and at low elevations it vignetted the
telescope.  This forced us to observe only one object in NGC 6528 per night,
and that only during the hour centered on culmination of the cluster.

Exposures were 1200 sec, and 3 such were obtained for star 3014 and star 3025,
while only two were obtained from 3046 and for 5422. The stars
were dithered by one or two arcsec along the length of the slit between
exposures.

b) Run of June 2000. Two exposures were obtained in NGC 6528 on the  nights of
June 2 and June 3, 2000 (i.e. 1 each night).  Each was 7200 sec long, in 1200
sec segments, with small spatial dithers along the slit between each segment.
Each exposure contained the RHB star 3025 and a second star, either 3014 or
5422.  Thus the total exposure for star 3025 was 14,400 sec. The HIRES setup
was identical for these two exposures. The S/N of these spectra, based
strictly on the count rate in the continuum near the center of the echelle
orders, is 65/pixel (130/4 pixel resolution element) for star 3025 and 95/4
pixel resolution element for the other two stars in NGC 6528.

The spectra from both runs were reduced using the suite of routines for analyzing echelle
spectra written by McCarthy (1988) within the Figaro image processing package
(Shortridge 1988).  The stellar data are flat fielded with quartz lamp
spectra, thereby removing most of the blaze profile, and the results are
normalized to unity by fitting a 10th-order polynomial to line-free regions of
the spectrum in each order.

\subsection{The Measurement of Equivalent Widths}

The spectra of the three stars observed in June, 2000 
(stars 3014, 3025 and 5422) are of high S/N, and
equivalent widths were measured directly from them; the August 1999
spectra were not used in the abundance analysis for these objects.
However, star 3046 was only observed in 1999.  The S/N of the 1999 spectra
is not very large (typically $\sim 30$).   For this
star only, we followed the same procedure successfully used in Paper I to
improve the reliability of measures of equivalent widths. We filtered the spectra
by convolving them with a Gaussian having a FWHM of 0.3~\AA\ (this reduced the
resolution of the spectra down to $\sim 15,000$ but enhanced the S/N per pixel
to about 60).   We also applied the above procedure to the spectra of the 3 
other stars of the 1999 run in order to derive a linear relationship
between the two sets of measured EWs, and thus to correct the 1999 EWs of
star 3046 measured using convolved spectra
to that of the higher precision June, 2000 spectra:

\begin{equation}
{\rm EW(2000)} = 0.90 {\rm EW(1999)} +12.5
\end{equation}
with the correlation coefficient $r=0.93$ and $\sigma$=19.7 from 361 lines.

Since the spectra of the program stars are very line rich, the following
procedure was adopted to measure reliable equivalent widths (EWs) on these
spectra. First, we corrected the continuum slightly by using a spline
interpolating function: appropriate ``continuum'' points were selected by
comparing the spectra of different stars, so that the fiducial continuum level
was set consistently for each star. This was done for all program stars. However, for star 3046, whose spectra
are at somewhat lower resolution, greater care is needed.
We then selected a small number of clean
lines, and used these lines to set a fiducial relation between the FWHM of
lines and their EWs. EWs for a larger number of lines were then measured by
using a special fitting routine that measures EWs using this relation between
FWHM and EWs, shown in Figure 3.  With this scheme, the FWHM, which is 
very sensitive to the presence of blending lines, was constrained when 
measuring the EWs of individual lines, resulting in much more stable measures. 

Note that for each line, the fitting
routine used an interactively selected portion of the line profile, so that
regions of the profile obviously disturbed by blending lines were not
considered. In this way, we are not using just the central pixel, but a 
portion of the line profile that is generally broader than the FHWM of the 
line (by itself broader than a resolution element), and we are fitting the 
central part of the line adopting a model of the line profile which is a 
Gaussian, whose FWHM is a (linear) function of the EW. 
As also noted by the referee, it is however possible that EWs are somewhat
overestimated. On the other side, since the continuum is likely to be somewhat
underestimated, if the two effects are not mutually compensating, we should
notice some trends of abundances as a function of the wavelength, and this is
not the case.

Apart from errors due to the continuum placement and to blends, random errors 
$\delta$EW in these EWs are:  $\delta$EW = $\frac{FWHM}{\sqrt{n} \cdot S/N}$,
where n is the number of points used in the fitting, and S/N is per pixel.
In our case n $\sim \frac{1.06 \cdot FWHM}{dx}$ (actually, somewhat larger, 
in general), 
where dx is the wavelength step; that is, we may write: 
$\delta$EW = $\frac{1.06 \cdot FWHM}{S/N*}$ where now S/N* is the S/N 
per FWHM element (generally larger than a resolution element).

Table 2\footnote{available only in electronic version.} lists the final values 
of the EWs, along with the adopted $gf$\ values.
Errors in the EWs measured by this procedure may be estimated by comparing the
EWs of different stars, since all four RHB stars have very similar atmospheric
parameters (see below). Through such comparisons, we obtain typical
rms scatters about the linear relationship between two stars of 9 m\AA. If we
assume that both sets of EWs have equal errors, we can estimate that typical
errors in EWs are 6 m\AA\ from the better quality spectra of June, 2000. These
errors are mainly due to uncertainties in the positioning of the fiducial
continuum. Errors in EWs measured on previous 1999 spectra (i.e. namely for
star 3046) are somewhat larger, between 10 and 12 m\AA.

\section {Abundance Analysis}

\subsection{Determination of Atmospheric Parameters}

The analysis of the CMD of NGC~6528 (Ortolani \etal\ 1992) revealed that the
interstellar reddening toward this cluster is quite large, somewhat uncertain,
and with strong variations even within small projected distances on the sky.
Infrared photometry by Cohen \& Sleeper (1995) confirmed the presence of
substantial reddening variations within this cluster. For this reason, it is
not possible to derive accurate atmospheric parameters from the observed
colours of stars in NGC 6528. Following the approach adopted in Paper I (where
we had a similar problem), effective temperatures for the
program stars were derived directly from the spectra, by forcing Fe~I lines of
different excitation to provide the same abundance (typically 90-100 Fe~I
lines were measured for each star, 70 for star 3046, with lower quality
spectra). The errors on the linear regression fits allow one to estimate the
(internal) errors in these temperatures: they are $\pm 85$~K (corresponding to
1$\sigma$ rms uncertainty of 0.017 dex/eV in the slope). As an example,
left panels of Figure~4 shows the run of abundances from individual
Fe~I lines with excitation potential for the RHB stars. Systematic errors
mainly depend on the set of model atmospheres used in the analysis (Kurucz,
1992: models with overshooting, for consistency with the analysis of Carretta
\& Gratton 1997). They are not easy to estimate, but we feel they are about
$\pm 100$~K.

Observation at slightly shorter wavelengths than done in Paper I allowed a
larger number of Fe~II lines to be measured (typically at least 4 good Fe~II
lines for each star): we were then able to estimate surface gravities $\log
g$\ from the equilibrium of ionization of Fe. Internal errors in these surface
gravities are $\pm 0.36$~dex (where we considered both the contributions due
to errors in EWs of individual lines, and those due to uncertainties in the
adopted $T_{\rm eff}$). Again systematic errors are mainly related to the
adopted model atmospheres, and to the assumption of LTE made throughout this
paper. (This seems a solid assumption for the stars we are currently
considering: see Gratton \etal\ 1999). The average gravity we derived
($2.16\pm 0.14$) is very close to that predicted by evolutionary models for
RHB stars in such a metal-rich cluster (about 2.3 dex, from the latest Padova
isochrones, Salasnich \etal\ 2000), supporting the temperature scale adopted
in the present analysis.

Microturbulent velocities $v_t$ were derived by eliminating any trend in the
derived abundances from Fe~I lines with expected equivalent widths for the
lines (following the approach of Magain 1984). Given the large number of Fe~I
lines measured, internal error bars in these $v_t$'s are small ($\pm
0.12$~km/s). As an example, the central panels of Figure~4 show the run of
abundances from individual Fe~I lines with EWs for the four program stars.

Of course, it is well known that in abundance analyses the derived values of
microturbulent velocity depend on the gf's one is using.
Our gf's are obtained by combining mainly two sources (see Carretta \& Gratton 
1997, henceforth CG97, for detailed references): (i) for all strong lines, 
and a few of the weak ones, we are using laboratory gf's from a compilation of 
data from the literature. 
Only gf's with errors $<$0.05 dex were considered; (ii) 
for the vast majority of weak lines (most of the lines measurable in the
spectra of NGC~6528 stars), we use solar gf's, obtained from an inverse
analysis whose zero point is set by the lines having laboratory gf's (and
whose abundance generally agree well with meteoritic values).

As noted also by the referee, there is some correlation between line strength
and excitation potential (\ep), in the sense that there is a 
generic tendency of
low-excitation Fe I lines to be stronger than those of high excitation. This
is shown by the curves-of-growth in the right-handed panels of Figure~4, where
open squares are used for lines with $\chi>3$ eV, while filled circles are for
lines with $\chi\leq 3$ eV.

To test the relevance of this problem in our analysis, we repeated it,
considering only those Fe I lines having $\chi> 3$ eV; we did not changed the
temperature since the excitation range included is now too small 
to allow fitting of $T_{\rm eff}$,
but we re-optimized the microturbulent velocity. 
When only lines with $\chi >3$ eV are considered, we are essentially using only
solar gf's; these lines are weak in the solar spectrum, so that these gf's
are virtually independent of details in the solar analysis.
The $v_t$ values we
derived from this subset of lines are smaller than the original ones by $0.07
\pm 0.06$ km/s, and the Fe abundances are larger by $0.03 \pm 0.03$ dex. Both
these values are barely significant and much lower than the other sources of
errors.

The final adopted atmospheric parameters are shown in Table~3.

An analysis of the influence of errors on the derived abundances is given in
Table~4. This table was obtained in the conventional way, specifically 
by comparing the abundances derived
for star 3025 with those derived by varying the atmospheric parameters, one at
a time, by the amount given in Table~4. As expected, larger effects arise
from uncertainties in $T_{\rm eff}$ (in particular for neutral species) and in
gravity (in particular for singly ionized species), whereas the overall metal
abundance and the microturbulent velocity only play a minor role. The last
column of Table~4 gives the quadratic sum of effects due to individual
parameters listed; this can be taken as an estimate of the total uncertainty
due to errors in atmospheric parameters. The uncertainty in O/Fe is by far the
largest entry in this column.

\subsection{Results}

One of our goals is to extend the calibration of the metallicity scale for
globular clusters of Carretta \& Gratton (1997) to the
metal-rich regime.  Therefore for the sake of consistency we adopt the same
atomic parameters (listed in Table~2) that they used, just as we did in 
Paper 1. Note that 
the only difference in the line lists used for NGC 6528 and NGC 6553 is in 
the slightly different wavelength region covered, due to the different 
instrumental configuration of HIRES used for the observations.
However, both lists are simply subsets of that used in CG97, securing the
required homogeneity.
In addition,
the same set of model atmospheres (Kurucz 1992 with convective overshooting),
code for abundance analysis, {\it etc.} previously adopted are also used in
the present work. Combining our previous work on NGC 6553 (Cohen \etal\ 1999)
with the present study doubles the sample of clusters with metal abundances
near the solar value, while retaining a highly homogeneous fine abundance
analysis for all clusters studied to date.

From the four RHB stars, we find that the mean [Fe/H] for NGC 6528 is $+0.07
\pm 0.01$\ dex, with a star-to-star scatter of $\sigma = 0.02$\ dex.
This is the first high dispersion analysis of a galactic globular cluster
in which an abundance greater than solar has been obtained.
Note however that the uncertainties are from statistics only. A fair 
estimate of the total error bar should include also systematic errors, that
are in general rather hard to quantify: including systematics, a conservative
budget could be about 0.1 dex, mostly due to errors related to the temperature
scale and to adopted model atmospheres.

The resulting abundances for each species in each star are listed in Table~5.
As in Paper I, all element ratios are computed with respect to Fe I, except
for O~I and Sc~II where we used abundances from Fe II to minimize the
uncertainties resulting from the choice of \teff. The rms dispersion in
abundance among the measured set of lines for each ion is given in parentheses
and the adopted solar abundances are shown in the last column of Table~5.
The results for star 3046 are from EWs converted to the same system of
the 3 other stars. Only abundances derived from at least 2 lines are shown for
star 3046.
Also, for comparison, we give in the last column of this Table the abundances
obtained by a similar analysis of the well known population~I star $\zeta$~Cyg
from Paper~I.
  
For oxygen and sodium (discussed in detail below), we give also the abundances
(from line analysis) including corrections for departures from LTE, following 
Gratton \etal\ (1999).

Abundances for Sc II, V I and Mn I include detailed corrections for the
quite large hyperfine structure of their lines (see Gratton 1989 and 
Gratton \& Sneden 1991 for references).

The Ca abundances were derived applying to each line used the collisional
damping parameter appropriate for that line (Smith \& Raggett 1981).

An estimate of Eu abundance was obtained by comparing the average spectrum 
of all the RHB stars in NGC 6528 observed in the region of the Eu II line at 
6645.11~\AA\ with synthetic spectra (Figure~5). From this comparison, evidence 
for a mild ([Eu/Fe] $\simeq 0.2$) overabundance of Eu is found.

\subsection{ Checks on the derived metal abundances }

A potential problem affecting the reliability of our results could be the
contamination of Fe I lines by blends. McWilliam and Rich (1994) demonstrated
that in bulge stars the effects of CN lines alone can be overwhelming at
metallicities whose mean is similar to the values derived here. We expect this
concern to be much less important in our spectra since they have a much higher
resolution than those used by McWilliam and Rich and the temperature of the 
stars is higher.
 However, in order to check
the impact of this problem, we considered more in detail the Fe I lines. For
each Fe I line included, we synthesized a spectral region of 3.2~\AA\ centered
on the line, using line lists extracted from Kurucz 1995 CD-ROMs (CD-ROM 23
for the atomic lines; and CD-ROM 15 for diatomic molecules; note that these
lists include lines due to CN as well as to other molecules) and the same
model atmospheres used for the program stars. (In this exercise, we assumed 
[C/Fe]$-0.5$ and [N/Fe]$+0.5$ since, as reminded by the second referee, RHB
stars have already experienced the phase along the red giant branch where C is
usually depleted and N enhanced, and their sum is constant.). The synthetic 
spectra were then convolved with a
Gaussian with a FWHM of 0.12~\AA, in order to match the resolution of our
spectra. We examined the profiles, and flagged all lines whose profiles are
in some way distorted by blends; we then measured the EWs of the lines on
these synthetic spectra using a Gaussian fitting routine (all parameters left
free; EWs obtained by this procedure are much more affected by the presence of
blends than those obtained using the procedure we applied for the program
stars).  

These EWs were then compared with those obtained from a spectral
synthesis where the line list only consisted of the line under scrutiny (using
this time a simple integration over the profile). Next we flagged all lines
where the two EWs differ more than 2~m\AA\ (note that the weakest Fe I lines
have EW$>40$~m\AA). Finally
we adopted as very clean lines those with no appreciable distortion in the profiles, and
for which the EWs are not changed by more than 2~m\AA\ by blends. 
We find a total of 53
very clean Fe I lines; such lines are marked with an asterisk in Table~2.
Average abundances obtained from this subset of lines are within 0.020 dex of those
derived from the original sample which in addition included lines of
somewhat poorer quality.  Furthermore the r.m.s. scatter of individual values are not
modified in a significant way.

This exercise suggests that (i) the procedure used to measure EWs (i.e. adopting
an average relation between EWs and FWHM to constrain the 
FWHM) allows us to derive reliable EWs even
for moderately blended lines as in the spectrum of star 3046; and (ii) 
that the largest source 
of errors in the EWs is the location of the continuum level rather than the
presence of blends.

Moreover, following the same procedure as in Paper~I, we checked our overall
[Fe/H] values by comparing our spectra with syntheses of a spectral region
around the Li doublet, which includes several weak or medium strength Fe lines.
These comparisons are shown in Figure~6, only for the 3 stars having new, high
quality spectra from the run of 2000. The remaining star, 3046, was tied on
the EW system defined from the 3 others as explained above.

The comparisons between the spectral syntheses and the observed
spectra for the RHB stars in NGC~6528
shown in Figure~6 support the abundances found by the analysis
of the equivalent widths. Lines computed with [Fe/H]=$-$0.13 and [Fe/H]=+0.27
are clearly too shallow and too strong with respect to the observations. 
Figure~6
also shows that no lithium is detectable in all our program stars. The
synthetic spectra in Figure~6 are computed with log n(Li) = $-$2\footnote{For
lithium abundances we used the usual notation: log~n(A) is the abundance (by
number) of the element A in the usual scale where log~n(H)=12; for all other
elements we use instead the notation [A/H], which is the logarithmic ratio of
the abundances of elements A and H in the star, minus the same quantity in the
Sun.}; however, since the Li line is blended with a Fe line, the upper limit
determined from our spectra is log n(Li)$<0$.

We also note that if we only use Fe I lines with
log(gf) - $\theta\chi \le -7.2$ (where $\theta$ = 5040/T), which corresponds
to the weakest lines that could be detected in the spectra taken
in the June 2000 run, we obtain [Fe/H]=+0.08 dex using 14 lines
for star 3025, [Fe/H]=+0.09 for star 5422 using 11 lines, and 
[Fe/H]=+0.06 for 12 lines in star 3014, which values are
indistinguishable from those obtained with the full set of Fe I lines.  
Abundances from such weak lines are
almost independent of the choice of microturbulent velocity. 
Although expected by the manner in which $v_t$ is set,
the close agreement
in the derived Fe abundance between the set of weak lines and the full set of
Fe I lines is reassuring.

\section{Discussion of Results}

The present results are summarized in Table~6, where we list also the mean
abundances for NGC 6553 (both from Paper I and from B99), and for 11 giants in
Baade's Window studied by 
McWilliam \& Rich (1994)\footnote{Rich \& McWilliam (2000) 
present a preliminary report suggesting that from their higher
S/N HIRES spectra they estimate that their previously
published Fe abundances for galactic bulge giants need to
be revised upwards by 0.1 to 0.2 dex for stars more metal rich than the
Sun.}, in order to provide a
deeper insight into our findings.  For NGC 6528 and our analysis of NGC 6553,
if there was only one line per star for a given ion, the value is given in
parentheses.

\subsection{Comparisons of Fe Abundance}

The most meaningful and immediate comparison is with our results for NGC 6553
from Paper I, since the analysis technique and data set are very consistent
and homogeneous. However, both the quality of the observational material and
the approach to the EW measurements present some differences.

Therefore, the first test we performed was a direct comparison of spectra of
stars in the two clusters. RHB stars are objects in a well defined evolutionary
phase.  Hence, we can expect their stellar stucture and parameters to be very similar
at similar metallicities.

This is evident from Figure~7, where the spectrum of star 71 in NGC 6553
(from Paper I) is compared to that of star 3025 of NGC 6528 (from the present work)
in the spectral region 6700-6740\AA. For a meaningful comparison, the spectra have
been degraded to the same resolution. This Figure shows that spectra of the two
stars are actually very similar, apart for some small differences in the
positioning of the continuum. We conclude that the two stars can
hardly be considered different or even distinguishable on the basis of their spectra.

This idea is strongly supported by Figure~8, where average EWs from stars of
NGC 6553 (Paper I) and NGC 6528 (present study) are compared. In order to make
the comparison more robust and less dependent on details of the continuum
tracing, we average only EWs of lines measured in at least 3 stars out of 5 in
NGC 6553 and in at least 2 stars out of the 3 with better quality spectra in
NGC 6528.

Again, this comparison is quite good and EWs are quite similar for a
hypothetical $average$ RHB star in the two clusters.

If the observed spectra are the same and 
the equivalent widths are very similar, then the difference of about 0.2 dex in
the average overall metallicity of these two globular clusters
([Fe/H]$=-0.16$ for NGC 6553 and
[Fe/H]$=+0.07$ dex in NGC 6528) must arise from the different
atmospheric parameters adopted in the two analysis. In fact, both here and in
Paper I, the atmospheric parameters adopted were derived directly from spectra, 
i.e. from EWs, that in turn were measured on spectra of different quality (better for
the NGC 6528 stars observed in the run of June 2000) and using different methods.

To verify this, we use Table~4 to estimate the changes in [Fe/H] as
due to differences between the mean atmospheric parameters used for RHB stars
in NGC 6553 (namely 4727/2.3/$-$0.13/1.82 for temperature, gravity, model
abundance and microturbulent velocity, respectively) and the mean set used for
NGC 6528 (namely 4620/2.21/+0.07/1.34). The resulting difference in [Fe/H] is
$-0.24$, in the sense that an $average$ star in NGC 6553 should be
more metal-poor than the $average$ one in NGC 6528.  The small difference in
adopted microturbulence of only 0.5 \kms\ gives rise to most of the 
abundance difference. 

Another experiment seems to confirm this finding. We used the set of average
EWs for NGC 6528 and repeated the analysis. We derive the values
4640/2.28/0.08/1.37 for $T_{\rm eff}$, $\log g$\, [A/H] and $v_t$, and obtain
[Fe/H]$=+0.08$ using the procedures discussed in
Section 3.1 of zeroing trends of abundance from single lines with \ep\ and
expected line strengths. With these values we also obtained a very good
ionization equilibrium.

If we now repeat the abundance analysis applying this set of atmospheric
parameters to the average EWs of NGC 6553, we obtain a solution where the
slopes of linear regressions of abundances $vs$ \ep\ and of expected line
strengths are well within the 1$\sigma$ rms uncertainty. The resulting
abundance for NGC 6553 is then [Fe/H]$=+0.04$, i.e. we can conclude that
an acceptable value for the overall metallicity of this cluster, as 
estimated from the Fe I abundance, is only 0.04 dex lower than that 
of NGC 6528. 

Finally, Figure~9 shows the parameter space $v_t$, $T_{\rm eff}$ over
the regime expected for RHB stars. Using the average EWs, we
show the contours which give  errors in slopes of linear
regressions of abundances $vs$ \ep\ and of expected line strengths within
the 1$\sigma$ rms uncertainty.
The loci for both NGC 6528
(solid line) and for NGC 6553 (dotted line) are elongated. This is simply
another, graphic representation of the coupling
that exists between $v_t$ and \teff\ 
for cool, metal-rich stars discussed above (Section 3.1).

The higher quality spectra taken in the second run for NGC 6528 do alleviate
this phenomenon, permitting 
high precision measurements of EWs of weak lines also of high excitation. 
This in turn allows us to better constrain the
region of this parameter space within which solutions of the abundance analysis are
acceptable.

These tests support the conclusion that within the error bars, the
overall metal abundance of the two clusters expressed in terms of [Fe/H] is
virtually the same. This is a firm result, since the number of measured Fe
lines is large.
Taking into account the overall uncertainty,  we
can be quite confident that NGC 6528 is a close ``twin" of NGC 6553, as far as
the overall metal abundance is concerned. Both these bulge clusters have a
metallicity slightly larger than solar. Extremely high abundances for NGC 6528
as well NGC 6553 can be safely ruled out, as confirmed also by their normal
integrated colours (see Figure 8 in Feltzig \& Gilmore 2000).

To be conservative, we will adopt hereinafter a value of metallicity for 
NGC 6553 that is the mean of that obtained in Paper I, and the one obtained 
with atmospheric parameters equal to those of NGC 6528. As error bar, we will
adopt one giving these two values as extremes of our confidence range. 
Hence, we will adopt for NGC 6553 a value of [Fe/H]$=-0.06 \pm 0.15$ dex.

\subsection{Comparison With NGC 6553 For Other Elements}

The pattern described by the element ratios appears to be rather
similar for the two clusters. The average abundances are generally in good
agreement, apart for the [Mg/Fe] ratio. However, as discussed in \S2.2, with
the present instrumental set up, we had no access to the Mg line at
8717.8~\AA, the clean line used in the analysis of RHB stars in NGC 6553. 
Instead we had to use two lines around 5600~\AA\ that are located in a more
crowded region and are saturated, hence not optimal for a good abundance
derivation. Even if the abundance from the line at 7657~\AA\ that we measured
for stars observed in the 2000 run generally confirms results from the other
lines, at present, and until further confirmation, we cannot be very confident
that the low Mg abundance found for NGC 6528 is real.

The overabundance of Si and of Ca seems to be well established in 
NGC~6528. Since, however, O abundances (discussed later) and Ti abundances seems
more likely solar, it is not clear that NGC 6528 presents the classical
overabundance of the $\alpha$-elements typical of more metal-poor clusters and
field stars.

On the other hand, Mn is rather underabundant in NGC~6528, and this result
is a signature of a small contribution to nucleosynthesis by SN Ia, suggesting
that the $\alpha$-elements overabundance might be the typical fossil 
record of a nucleosynthesis history heavily dominated by massive stars.
Some support to this scenario could be given by the mild overabundance of
Eu found from the combined spectrum of the RHB stars in NGC 6528, since Eu is 
known to
be a n-capture element predominantly produced by r-processes. Note however that
the ratio [Ba/Eu] is almost solar, suggesting that maybe we are seeing also the
contribution to nucleosynthesis from intermediate-mass stars.

A comparison of our results for NGC 6553 with those of B99 was given in
section 7.1 of Paper I. Here, we note only that the choice of solar and
stellar atmosphere model adopted in B99 (hence the solar abundances) results
in Fe abundances that are 0.1 to 0.15 dex lower than those of the present
scale. This by itself decreases the differences in the element ratios
presented in Table~6. In other words, had the comparison for NGC~6553 been
made, instead, in terms of [element/H], the pattern of many elements would
have been more similar when our analysis of Paper I was compared to that of
B99.

\subsection{O and Na Abundances}

The slightly different instrumental configuration of HIRES adopted for the
observations of the RHB stars in NGC 6528 allowed the use of the weak Na
doublet at 6154, 6160~\AA\ as well as the infrared permitted O I triplet, and,
for three of the four stars, the forbidden 6300~\AA\ O I line as well.

To take into account departures from the adopted assumption of LTE,
appropriate corrections were applied to O and Na abundances following the
prescriptions of Gratton \etal\ (1999). Non-LTE abundances from the line
analysis are also listed in Table~5. Note however, that the corrections are
rather small, and do not exceed a few hundredths of a dex in all cases.

We tested the abundances we derived for Na and O using synthetic spectra of
the 6154-60 Na~I doublet, and of the [O~I] line at 6300 \AA.
The spectral synthesis were computed assuming LTE.
Comparisons with
spectral synthesis for these two spectral regions are shown in Figures 10 and 11
respectively. 

From the comparisons in Figure~10 we found that Na is overabundant by about
+0.35 to +0.5 dex in stars 3025, 5422, and 3014 (respectively, we estimated
[Na/Fe]= +0.5, +0.35 and +0.4 dex). These results are consistent with that
deduced from the EWs.

The comparison between observed and synthesized spectra for the region
including the [OI] line at 6300.3 \AA\ is more difficult for three reasons.
First, there are several telluric lines in this region whose
locations are marked with a T in Figure~11. Due to the
high $v_r$ of NGC~6528, the
wavelength of the [OI] line in the observed spectra of NGC~6528 is
free of contamination by telluric lines; they
should not affect our results. Second, as mentioned in Section 3.1, O 
abundances are quite sensitive to uncertainties in the atmospheric parameters.
Third, there is some coupling between C and O
abundances, due to formation of CO in the atmospheres of these cool,
metal-rich K-giants. We estimated that derived O abundances increase by 0.06
dex if the C abundances are raised by 0.2 dex. Unfortunately, we could not
determine C abundances from our spectra, since EWs of C I lines measured in
the red are not reliable; we can only fix the overall strength
of CN lines. We matched adequately CN lines by assuming [C/Fe]=$-$0.45 and
[N/Fe]=0 when [O/Fe]=$-$0.07; [C/Fe]=$-$0.3 and [N/Fe]=0 when [O/Fe]=0.13; and
[C/Fe]=$-$0.18 and [N/Fe]=0 when [O/Fe]=+0.33. These are the values adopted
in the syntheses shown in Figure 11. 

Figure 11 shows that with the adopted C and N abundances, a good fit of the
[OI] line is obtained for [O/Fe]$\sim 0$, in reasonable agreement with the
results given by the analysis of the equivalent widths (where also other
O lines are taken into account). An O excess of 0.3 dex seems excluded,
unless the C abundance is not much larger than solar, which seems unlikely.

We conclude that the program stars in NGC~6528 have solar oxygen abundances,
but a distinct excess of Na ([Na/Fe]$\sim 0.4$).

\subsection{Metal Abundances: Calibration of Other Indices}

Our need, after the completion of our analysis of NGC 6553 of Paper I, to have
an analysis for stars in at least one more metal-rich cluster is apparent if
one examines Figure~12, where the metallicities obtained from high resolution
spectra by CG97, plus those for NGC 6553 and NGC 6528 (this study)
are plotted against the parameter W(CaII) determined by Rutledge \etal\ (1997)
from the measured strength of the infrared Ca triplet in individual globular
cluster giants.

For reasons explained in Section 4.1, we believe that the
correct [Fe/H] values to adopt for these two clusters are +0.07 and 
$-$0.06 dex for NGC 6528 and NGC 6553, respectively.

When NGC 6528, which has the largest value of any galactic
globular cluster observed to date on the Rutledge \etal\ ranking
scale (W(CaII)=5.41), is added, the conclusion proposed in Paper I is still
firmly supported. Specifically, the two scales seem to be linearly
correlated\footnote{Note that in Figure~12 a dashed line indicates the linear
fit that one would obtain using all clusters, just for purpose of comparison}
over the range from [Fe/H]$=-2.2$ to [Fe/H]$=-0.6$ (i.e. up to not so
extremely metal rich clusters). At higher metallicities, however, where the
CaII index is known to lose its sensitivity to metal abundance (see Armandroff
\& Da Costa 1991), the linear correlation seems to break down.

Note that clusters like Pal 12 or Rup 106, known to have anomalously low 
[$\alpha$/Fe] ratios, are not included in the sample.

With our new results we can now better assess the relationship between the two
scales: in fact, NGC 6528 falls very near NGC 6553, in the [Fe/H]--W(CaII)
plane, and strongly constrains the position of metal-rich globular clusters
for the metallicity calibration in terms of this low resolution indicator. This
is not unexpected, since in Section 4.1 we showed how RHB stars of these 2
clusters are similar. We find that to bring the values from Rutledge \etal\
onto a homogeneous metallicity scale which is based completely on high
dispersion spectroscopy, one has to adopt a second order polynomial relation:

\begin{equation}
{\rm [Fe/H]_{us}} = -2.08 -0.04 {\rm W(CaII)} + 0.078 {\rm W^2(CaII)}
\end{equation}
with the correlation coefficient $r=0.98$ and $\sigma$=0.12 dex for 22 clusters.
The second order term is significant at a confidence level
exceeding 95\%; inclusion of this term yields $\chi^2 = 25.8$
for the sample of 22 globular clusters in common.  This is
a reduction in $\chi^2$ of
a factor of 1.6 compared to a linear fit.
Notice that similar results are obtained even if we omit entirely NGC 6553,
so that the larger error bar of the revised [Fe/H] value for this cluster
does not affect much our calibration.

This relation, shown in Figure~12, allows us to derive directly from the Ca
index of a globular cluster its metallicity on the Carretta \& Gratton scale, 
as extended
here and in Paper I. The range of application should be restricted to that of
calibrating clusters, namely from W(CaII) $\sim 1.5$ (NGC 7078,NGC 4590) to
W(CaII) $\sim 5.4$ (NGC 6528).
Moreover, using data from the compilation of Carney (1996), we checked that
the differences between the W(CaII) values as observed by Rutledge \etal\ and 
the W(CaII) values predicted from our eq. (2) are constant as a function of [Ca/Fe].
We can then conclude that, since the [Ca/Fe] ratio holds rather constant for all
clusters in the sample, we are actually calibrating an index that is tied to the
temperatures and luminosities of stars along the RGB, which in turn depend on 
the cluster metal abundance.

Adding a second well analyzed high metallicity cluster is also very
useful in intercomparing the CG97 scale and the widely used
scale of Zinn \& West (1984) (based on integrated light indices). We are now 
able to
derive a relation to bring their values on the new scale entirely based on
high dispersion analyses, without using any uncertain extrapolations.
As in CG97, we prefer to use directly final metallicities from Zinn \& West
(1984), with the update of Armandroff \& Zinn (1988; collectively ZW), due to
the variety of indicators used in the analysis of Zinn \& West (1984).

When NGC 6528 and NGC 6553 are added to the other 24 calibrating clusters, the
quadratic relation previously found by CG97 is no longer the most appropriate,
as one can see in Figure~13, where mean metallicities from the CG97 scale as
augmented here are compared with metallicities on the ZW scale.
 
In this case a cubic polynomial is the best relation to transform the
ZW scale to the new high dispersion spectroscopic scale: 
\begin{equation}
{\rm [Fe/H]_{CG}} = + 0.61 +3.04 {\rm [Fe/H]_{ZW}} +1.981 
{\rm [Fe/H]_{ZW}}^2 +0.532 {\rm [Fe/H]_{ZW}}^3
\end{equation}
with $\sigma$=0.10 dex and a correlation coefficient $r=0.99$ for 26 clusters.

When considering the adopted error bar of 0.15 dex for the revised value of
[Fe/H] of NGC 6553, 
the second and third order terms are significant at the
99\% confidence level.  

The value of $\chi^2$ decreases by more than
a factor of two in going from a linear fit to a third order fit,
e.g. from 392 to 144 for the comparison of our abundances
with those of Zinn \& West (1984) for the 26 clusters (25 without
NGC 6553) in the sample with high dispersion abundance analyses.

The present work extends the range of application of this transformation by
about 0.3 dex toward high metallicity, with respect to eq. 7 in CG97. Using
the above relation now allows one to transform the ZW metallicities onto our
high dispersion scale in the range $-2.24<$[Fe/H]$_{\rm ZW}<+0.12$.

\section{Conclusions}

We present for the first time an abundance analysis for stars in the
metal-rich bulge cluster NGC 6528 based on high resolution spectra of S/N high
enough to allow a reliable fine abundance analysis.

We observed 4 red HB stars, from which we found a mean [Fe/H]$= +0.07 \pm
0.01$\ dex, with a star-to-star scatter of $\sigma = 0.02$ (not including
systematic effects).

This metallicity is slightly larger than the mean abundance in the galactic
bulge found by McWilliam \& Rich (1994), and that found in a previous paper
for NGC 6553. However, we found that the spectra of stars in NGC~6553
and NGC~6528, and the EWs we measured on them, are very similar to each other. 
The slightly different metal
abundance is the result primarily of adoption of slightly lower 
microturbulent velocity in the analysis of NGC~6528 stars
and secondarily of the adoption of slightly lower
temperatures for the  NGC~6528 stars. 

Note that the present analysis is based on higher quality spectra, so that the 
atmospheric parameters here used for NGC 6528 are more firmly established than
those used in Paper I for NGC 6553.

If the same
atmospheric parameters are adopted for both samples, nearly equal metal
abundances are derived for the two clusters. 
We then propose to revise upward the metal abundance of NGC~6553
to [Fe/H]$=-0.06 \pm 0.15$, where the error bar encompasses both the original
abundance for NGC 6553 we derived in Paper I, and the value we obtained 
assuming in the analysis  atmospheric parameters identical to those we used 
for stars in NGC 6528.

These results imply that metal-rich globular clusters may reach or even exceed
the mean abundance of the galactic bulge found by McWilliam \& Rich, but are
found to fall within the spread of their distribution.

The relative abundance for the best determined $\alpha$-process elements (Ca)
indicates an excess of $\alpha$ process elements of about a factor of two and
a global pattern of abundances similar to that of NGC 6553 and the bulge field
stars. 
Moreover, Mn in NGC~6528 clearly shows an underabundance typical of a
quite small contribution to nucleosynthesis by yields from SN Ia. When coupled 
with the overabundances of Si and Ca, our results strongly suggest that 
these bulge globular clusters seem to have experienced a history of chemical
enrichment essentially identical to that typical of bulge field stars, 
probably under conditions of enrichment by type II Supernovae at early
epochs.

We found a solar O abundance, while there is a clear excess of Na 
([Na/Fe]$\sim 0.4$). However the sample is not large enough to 
understand if this abundance
pattern is characteristics of all stars in this cluster, or rather stems from
peculiarities of the program stars or their evolutionary state.

Our new data for NGC 6553 (Paper I) and NGC 6528 allow us to re-calibrate the
widely used metallicity scales of Zinn \& West (1984) and of Rutledge \etal\
(1997), both based on low dispersion or integrated indices, onto a scale fully
based on high dispersion spectroscopy. We give the functions required to
transform W(CaII) and [Fe/H]$_{ZW}$ values into our updated scale, superseding
the previous calibration by Carretta \& Gratton (1997).

\appendix
\section{The Reddening of NGC 6528}

Like other bulge clusters heavily reddened and/or affected by differential
reddening, the value of E(B-V) for NGC 6528 in the literature ranges over more
than 0.2 mag, e.g. 0.46 mag (Richtler \etal\ 1998) from $V,I$ photometry, 0.55
mag (Ortolani \etal\ 1992 and Zinn 1980), and 0.73 mag from cluster integrated
colours (Reed \etal\ 1988).

Since an underestimate of 0.05 mag in the value of E(B-V) translates into an
underestimate of about 0.2 dex in metal abundance when temperatures are
derived from colours (Cohen 1983), this issue deserve special care. Even
methods such as that of Sarajedini (1994), which uses fits to the shapes of
unreddened giant branches to derive metallicity and reddening simultaneously,
are unreliable for very metal-rich clusters, since the most metal-rich of his
calibrating clusters is 47 Tuc ([Fe/H]$=-0.70$).

With the present analysis, we have a spectroscopic determination of the
temperatures of four RHB stars that are derived only from the excitation of
many Fe I lines.  These estimates are therefore reddening-free and provide an
independent route to the value of E(B-V) in NGC~6528.

As a starting point we assumed a mean value of $T_{\rm eff} \sim$ 4610 K as
representative of the spectroscopic temperatures of our RHB stars (see
Table~3). This temperature, with the mean metallicity and gravity from
Table~3, corresponds to a value (V-K)$_0 = 2.53$ based on the 
colour--$T_{\rm eff}$
calibration of Gratton, Carretta \& Castelli (1997)\footnote{We don't have
B-V colours of the program stars and we lack good I photometry for
bright stars which were used to derive the zero point correction to our 
colour--$T_{\rm eff}$ calibrations. Hence, the V-I and B-V colours do not add 
useful information to what we are deriving from V-K colours.}. On the other 
hand, from
the V,K photometry of Montegriffo (1999) the mean V-K colour for the 4 program
stars is 3.88; so that, assuming a standard relation E(V-K)=2.7 E(B-V) (e.g.
Cardelli \etal\ 1989), we deduce a value of E(B-V) = 0.50 mag as an
independent estimate for the reddening of NGC 6528. Repeating the above
exercise using the $individual$ colours and parameters of each star,  we could
not isolate any significant difference in the resulting values of E(B-V).

This value is somewhat lower than the one listed in the compilation by Harris
(1996), as well most of the values quoted above, although only slightly
smaller that that found in the field just N and W of the cluster by Stanek
(1996). (Stanek's value in the field just S of the cluster is close to that
of Harris.)
We believe that this arises
because of a selection bias.  In a metal-rich globular cluster with
differential reddening, the RHB shape in a CMD changes from the normal short
horizontal one characteristic of a constant reddening to an extended
distribution sloping towards fainter magnitudes for the redder stars.  Such a
RHB is clearly shown in the case of NGC~6553 in Figure 2b of Ortolani \etal\
(1990) and NGC 6528 (in Figure~2). Our sample of RHB stars selected for high
dispersion spectroscopy is biased towards the brighter stars, and hence the
less reddened ones. Table~3 shows, however, that the program stars have a
total range in $T_{\rm eff}$ of 70 K (well within the adopted uncertainties),
even though they cover a rather large range in the observed reddened CMD.

NGC 6528 is known to be affected by differential reddening. Cohen \& Sleeper
(1995) establish that the range of E(B-V) is about 0.25 mag.  Consistent with
the suggestion of differential reddening, it is interesting to note that three
of the program stars are located in the same region of NGC 6528. For
differential reddening of the magnitude found in NGC~6528, this selection
effect far outweighs any tendency toward picking RHB stars that are slightly
evolved off the zero age HB.

Irrespective of the origin of this effect and of the mean value of E(B-V) for
NGC~6528, we stress here that the determination of temperatures by line
excitation is a purely spectroscopic, reddening-free method and that this is
the method used for the four RHB stars we observed.

As noted by the referee, one may wonder how sensitive this method could be to
errors in gf values. In previous section, we performed a test
aimed to understand if our values for the microturbulent velocity were biased
due to the correlation existing between line strength and excitation potential,
because lines of low excitation are all strong. Hence we compared the values
for the $v_t$ derived from all lines with those we
derive using only high excitation lines. This second value is nearly
independent of the assumed temperature, because all lines have nearly the same
excitation potential.

We checked the zero-point of this calibration using field red clump stars 
(Carretta \etal, Cohen \etal\ in preparation) with atmospheric parameters 
similar to those of stars in NGC 6528 and with good parallaxes from 
Hipparcos. These stars are sufficiently close that it is reasonable to assume
that the reddening is negligible.

\acknowledgements 
We are grateful to Paolo Montegriffo for providing the HST and IRAC photometry 
of NGC 6528 prior to publication.
We warmly thank Bernardo Salasnich 
for constructing for us the isochrones from the latest Padova models in advance
of publication.  JGC and BBB are grateful for partial support from 
NSF grant AST--9819614.
The entire Keck/HIRES user community owes a huge debt
to Jerry Nelson, Gerry Smith, Steve Vogt, and many other people who have
worked to make the Keck Telescope and HIRES a reality and to
operate and maintain the Keck Observatory.  We are grateful
to the W. M. Keck Foundation, and particularly its late president,
Howard Keck, for the vision to fund the construction of the W. M. Keck
Observatory. This research has made use of
the SIMBAD data base, operated at CDS, Strasbourg, France.

\clearpage
%
%
\begin{deluxetable}{lrrrrrrrr}
\tablenum{1}
\tablewidth{0pt}
\scriptsize
\tablecaption{The Sample of Stars Observed in NGC 6528}
\label{tab1}
\tablehead{\colhead{ID} & \colhead{$V$\tablenotemark{a}} &
\colhead{$I$\tablenotemark{a}} &
\colhead{$J$\tablenotemark{a}} &
\colhead{$K$\tablenotemark{a}} & \colhead{$v_r$\tablenotemark{b}} &
\colhead{Date of Obs.} & \colhead{Exp. Time} & \colhead{Coord.} \nl
\colhead{} & \colhead{(mag)} & \colhead{(mag)} & \colhead{(mag)} &
\colhead{(mag)} & \colhead{(\kms)} & \colhead{} & \colhead{(sec)}
&\colhead{(J2000)}
}
\startdata
Red HB Stars \nl
5422  & 16.95 & 15.34 & 14.11 & 13.14 &  $206.8 \pm 1.0 \pm2.5$ & 990817 &
2 X 1200  & 18 04 47.58 $-$30 03 30 \nl
      &       &       &       &       &  $206.4 \pm0.7 \pm1.5$
& 000603 &    6 X 1200  & \nl
3025  & 17.14 & 15.43 & 14.21 & 13.22 &  $218.1 \pm1.0 \pm2.5$ & 990817 &
3 X 1200  & 18 04 47.62  $-$30 03 36 \nl
      &       &       &       &       &  $216.8 \pm0.7 \pm1.5$
& 000603 &   12 X 1200  & \nl
3014  & 17.09 & 15.38 & 14.21 & 13.21 &  $212.0 \pm1.0 \pm2.5$ & 990815 &
3 X 1200  & 18 04 47.38  $-$30 03 43 \nl
      &       &       &       &       &  $205.7 \pm0.7 \pm1.5$
& 000603 &    6 X 1200  & \nl
3046  & 17.25 & 15.50 & 14.12 & 13.34 &  $207.3 \pm1.0 \pm2.5$ & 990818 &
2 X 1200  & 18 04 51.52  $-$30 03 20 \nl
5425  & 17.09 & 15.35 & 14.25 & 13.32 &  $209.2 \pm0.8 \pm2.5$ & 000603 &
$<2500$  & 18 04 47.62  $-$30 03 25 \nl
3032\tablenotemark{c} & 16.69 & 14.71 & 13.58 & 12.34 &  
$211.2 \pm0.8 \pm2.5$ & 000603 & $<2500$  & 18 04 47.34  $-$30 03 33 \nl
\enddata
\tablenotetext{a}{From Montegriffo (1999; private communication)}
\tablenotetext{b}{The internal error in $v_r$ is followed by the systematic
error.}
\tablenotetext{c}{This star is a red giant in NGC~6528, not a RHB star.}
\end{deluxetable}

%
%
\begin{deluxetable}{lrrrrrrrrrrrrrrrrrr}
\tablenum{2}
\tablewidth{0pt}
\tabcolsep 2pt 
\scriptsize
\tablecaption{Equivalent Widths For 4 Red HB Stars in 
NGC 6528\tablenotemark{a}\tablenotemark{b}}
\label{tab2}
\tablehead{
\colhead{Ion} & \colhead{~~~~$\lambda$ (\AA)}	&\colhead{$\chi$ (eV)}	
&$\log(gf)$
&\mc{5422}	&\mc{3025} 		&\mc{3014}	&\mc{3046} \nl 	
  & & & & \mc{(m\AA)} & \mc{(m\AA)} & \mc{(m\AA)} & \mc{(m\AA)} 
}
\startdata
O I  & 6300.31 &  0.00 &  $-$9.75 &&  40.0 &&  39.0 &&  37.0 &&       \nl
O I  & 6363.79 &  0.02 & $-$10.25 &&   9.0 &&  15.0 &&  13.0 &&       \nl
O I  & 7771.95 &  9.11 &     0.33 &&  40.0 &&  36.0 &&  35.0 &&  54.8 \nl
O I  & 7775.40 &  9.11 &  $-$0.03 &&  30.0 &&  31.0 &&       &&       \nl
Na I & 5682.65 &  2.10 &  $-$0.67 && 182.9 && 194.4 && 200.7 &&       \nl
Na I & 5688.22 &  2.10 &  $-$0.37 && 184.8 && 187.1 && 208.6 &&       \nl
Na I & 6154.23 &  2.10 &  $-$1.57 && 114.5 && 125.3 && 116.5 && 101.6 \nl
Na I & 6160.75 &  2.10 &  $-$1.26 && 140.8 && 137.2 && 142.9 && 119.6 \nl
Mg I & 5528.42 &  4.34 &  $-$0.62 && 270.1 && 264.5 && 276.3 &&       \nl
Mg I & 5711.09 &  4.34 &  $-$1.83 && 156.6 && 157.5 && 155.0 && 132.0 \nl
Mg I & 7657.60 &  5.11 &  $-$1.28 && 124.5 && 115.7 && 165.9 &&       \nl
Si I & 5645.62 &  4.93 &  $-$2.14 &&  84.4 &&       &&  96.2 &&       \nl
Si I & 5665.56 &  4.92 &  $-$2.04 &&  93.5 &&  80.3 &&  99.5 &&  78.2 \nl
Si I & 5684.49 &  4.95 &  $-$1.65 &&  75.8 &&  95.1 && 112.5 &&       \nl
Si I & 5690.43 &  4.93 &  $-$1.87 &&  75.7 &&  95.3 &&  78.8 &&  90.2 \nl
Si I & 5701.11 &  4.93 &  $-$2.05 &&  61.5 &&  73.1 &&       &&       \nl
Si I & 5708.41 &  4.95 &  $-$1.47 && 129.8 &&       &&       &&       \nl
Si I & 5772.15 &  5.08 &  $-$1.75 &&  95.1 && 103.0 &&  83.9 &&  87.2 \nl
Si I & 5793.08 &  4.93 &  $-$2.06 &&  91.8 &&  99.2 &&  94.8 &&       \nl
Si I & 5948.55 &  5.08 &  $-$1.23 && 127.4 && 117.2 && 133.9 && 108.7 \nl
Si I & 6125.03 &  5.61 &  $-$1.57 &&  69.6 &&  59.9 &&  60.7 &&  54.5 \nl
Si I & 6145.02 &  5.61 &  $-$1.44 &&  51.6 &&  56.2 &&  59.2 &&  51.9 \nl
Si I & 6848.57 &  5.86 &  $-$1.75 &&  50.5 &&  47.2 &&  42.7 &&       \nl
Si I & 6976.50 &  5.95 &  $-$1.17 &&       &&  83.8 &&       &&       \nl
Si I & 7003.58 &  5.96 &  $-$0.87 &&       &&  63.9 &&  70.2 &&       \nl
Si I & 7034.90 &  5.87 &  $-$0.88 &&  93.7 &&  86.0 && 110.2 &&       \nl
Si I & 7226.20 &  5.61 &  $-$1.51 &&       &&       &&       &&  60.6 \nl
Si I & 7405.79 &  5.61 &  $-$0.82 && 111.4 && 120.7 && 126.6 &&       \nl
Ca I & 5590.13 &  2.51 &  $-$0.57 && 128.1 &&       && 150.9 &&       \nl
Ca I & 5594.47 &  2.51 &     0.10 && 224.5 && 215.0 && 213.1 &&       \nl
Ca I & 5857.46 &  2.93 &     0.24 && 182.9 && 178.4 && 208.0 && 184.3 \nl
Ca I & 5867.57 &  2.93 &  $-$1.49 &&  66.7 &&  75.2 &&  86.8 &&  62.8 \nl
Ca I & 6161.30 &  2.52 &  $-$1.27 && 122.1 && 129.0 && 114.0 &&       \nl
Ca I & 6166.44 &  2.52 &  $-$1.14 && 107.4 && 120.0 && 122.5 && 119.7 \nl
Ca I & 6169.04 &  2.52 &  $-$0.80 && 145.0 && 142.9 && 150.7 &&       \nl
Ca I & 6169.56 &  2.52 &  $-$0.48 && 164.5 && 164.3 && 160.9 &&       \nl
Ca I & 6439.08 &  2.52 &     0.39 &&       &&       &&       && 234.9 \nl
Ca I & 6462.57 &  2.52 &     0.26 && 299.3 && 282.5 && 299.7 &&       \nl
Ca I & 6471.67 &  2.52 &  $-$0.69 && 152.9 && 153.1 && 162.8 && 128.5 \nl
Ca I & 6493.79 &  2.52 &  $-$0.11 && 174.9 && 185.3 && 197.4 && 193.0 \nl
Ca I & 6499.65 &  2.52 &  $-$0.82 && 130.7 && 144.6 && 147.7 && 145.7 \nl
Ca I & 6572.80 &  0.00 &  $-$4.32 &&       &&       &&       && 139.7 \nl
Ca I & 6717.69 &  2.71 &  $-$0.52 && 197.9 &&       &&       &&       \nl
Sc II& 5526.82 &  1.77 &     0.19 && 140.0 && 127.2 && 128.6 &&       \nl
Sc II& 5640.99 &  1.50 &  $-$0.86 && 103.4 && 101.5 &&       &&       \nl
Sc II& 5657.88 &  1.51 &  $-$0.29 && 119.4 && 121.5 && 120.6 &&       \nl
Sc II& 5667.15 &  1.50 &  $-$1.11 &&  80.8 &&  81.5 &&       &&       \nl
Sc II& 5669.04 &  1.50 &  $-$1.00 &&  86.8 &&  81.0 &&  92.3 &&  64.8 \nl
Sc II& 5684.20 &  1.51 &  $-$0.92 &&       &&  82.6 && 102.4 &&       \nl
Sc II& 6245.62 &  1.51 &  $-$1.05 &&  88.1 &&  89.6 &&  98.1 &&  79.2 \nl
Sc II& 6604.60 &  1.36 &  $-$1.14 &&  88.7 &&  92.8 && 100.9 &&  89.8 \nl
Ti I & 5490.16 &  1.46 &  $-$0.93 &&  97.1 && 101.3 &&  87.7 &&       \nl
Ti I & 5503.90 &  2.58 &  $-$0.19 &&  81.9 &&       &&       &&       \nl
Ti I & 5662.16 &  2.32 &  $-$0.11 &&       &&  87.8 && 102.1 &&       \nl
Ti I & 5689.48 &  2.30 &  $-$0.47 &&       &&       &&  57.4 &&  60.4 \nl
Ti I & 5866.46 &  1.07 &  $-$0.84 && 144.6 && 139.3 && 160.9 && 123.3 \nl
Ti I & 5922.12 &  1.05 &  $-$1.47 && 102.5 && 111.7 && 106.3 &&  90.8 \nl
Ti I & 5978.55 &  1.87 &  $-$0.58 && 105.5 &&  96.3 && 108.0 &&       \nl
Ti I & 6091.18 &  2.27 &  $-$0.42 &&  68.2 &&  69.9 &&       &&       \nl
Ti I & 6126.22 &  1.07 &  $-$1.42 &&  93.1 && 104.9 && 107.9 &&  87.9 \nl
Ti I & 6258.11 &  1.44 &  $-$0.36 && 122.7 && 131.7 && 134.7 &&       \nl
Ti I & 6261.11 &  1.43 &  $-$0.48 && 166.8 &&       && 157.2 &&       \nl
Ti I & 6554.24 &  1.44 &  $-$1.22 &&       &&       &&       && 109.2 \nl
Ti I & 7251.72 &  1.43 &  $-$0.84 && 122.7 && 126.3 && 125.2 && 148.7 \nl
V I  & 5627.64 &  1.08 &  $-$0.37 &&       &&       &&       && 103.8 \nl
V I  & 5670.86 &  1.08 &  $-$0.42 && 116.1 && 110.7 && 105.0 &&  87.1 \nl
V I  & 5703.59 &  1.05 &  $-$0.21 && 111.3 && 113.8 && 113.9 &&       \nl
V I  & 5727.06 &  1.08 &  $-$0.01 &&       &&       &&       && 142.6 \nl
V I  & 6081.45 &  1.05 &  $-$0.58 &&  98.9 &&  95.2 && 116.8 &&       \nl
V I  & 6090.22 &  1.08 &  $-$0.06 && 120.8 && 115.0 && 130.6 &&       \nl
V I  & 6119.53 &  1.06 &  $-$0.32 &&       &&       &&       && 107.4 \nl
V I  & 6199.19 &  0.29 &  $-$1.28 && 141.5 && 134.0 && 148.7 &&       \nl
V I  & 6216.36 &  0.28 &  $-$1.29 &&       &&       &&       && 138.5 \nl
V I  & 6243.11 &  0.30 &  $-$0.98 &&       &&       && 180.6 &&       \nl
V I  & 6251.83 &  0.29 &  $-$1.34 && 116.2 && 117.2 && 116.9 && 113.6 \nl
Cr I & 5409.80 &  1.03 &  $-$0.71 &&       && 268.1 &&       &&       \nl
Cr I & 5702.33 &  3.45 &  $-$0.68 &&       &&  76.6 &&  76.1 &&       \nl
Cr I & 5781.19 &  3.32 &  $-$0.88 &&  79.4 &&  64.1 &&  55.6 &&       \nl
Cr I & 5781.76 &  3.32 &  $-$0.75 &&  81.1 &&  67.5 &&  69.1 &&       \nl
Cr I & 5783.07 &  3.32 &  $-$0.40 &&       &&       &&  59.6 &&  53.6 \nl
Cr I & 5783.87 &  3.32 &  $-$0.30 &&       &&  96.0 &&  89.5 &&  97.8 \nl
Cr I & 5784.98 &  3.32 &  $-$0.38 &&  89.6 &&  80.5 &&  96.9 &&       \nl
Cr I & 5787.93 &  3.32 &  $-$0.08 &&  97.5 &&  80.9 &&  82.8 &&       \nl
Cr I & 6330.10 &  0.94 &  $-$2.87 &&       &&       &&       && 115.1 \nl
Cr I & 6882.50 &  3.44 &  $-$0.38 &&       &&  65.8 &&       &&       \nl
Cr I & 6883.07 &  3.44 &  $-$0.42 &&  65.6 &&  64.2 &&  71.9 &&       \nl
Cr I & 6882.52 &  3.44 &  $-$0.38 &&       &&  65.8 &&       &&       \nl
Cr I & 6883.06 &  3.44 &  $-$0.42 &&  65.6 &&  64.2 &&  71.9 &&       \nl
Cr I & 6979.80 &  3.46 &  $-$0.22 &&  76.0 &&  86.8 &&  77.0 &&       \nl
Cr I & 6980.94 &  3.46 &  $-$1.09 &&  40.3 &&       &&       &&       \nl
Cr I & 7400.19 &  2.90 &  $-$0.11 && 132.6 && 151.6 && 151.8 && 162.5 \nl
Cr II& 5502.09 &  4.17 &  $-$1.96 &&  46.0 &&  46.1 &&       &&       \nl
Cr II& 5508.63 &  4.15 &  $-$2.07 &&  57.7 &&  38.7 &&  53.3 &&       \nl
Mn I & 5394.67 &  0.00 &  $-$3.50 &&       && 199.7 &&       &&       \nl
Mn I & 5420.37 &  2.14 &  $-$1.46 &&       && 192.9 &&       &&       \nl
Mn I & 5432.55 &  0.00 &  $-$3.80 && 176.1 && 188.4 &&       &&       \nl
Mn I & 6013.50 &  3.07 &  $-$0.25 &&       &&       &&       && 146.1 \nl
Mn I & 6016.65 &  3.07 &  $-$0.09 &&       &&       &&       && 146.2 \nl
Mn I & 6021.80 &  3.08 &     0.03 && 183.7 && 133.7 && 165.4 && 150.7 \nl
Mn I & 7302.85 &  4.43 &  $-$0.37 &&  59.5 &&  58.8 &&  52.0 &&       \nl
Fe I & 5386.34*&  4.15 &  $-$1.74 &&       &&  70.7 &&       &&       \nl
Fe I & 5389.49 &  4.41 &  $-$0.57 &&       && 105.5 &&       &&       \nl
Fe I & 5395.22 &  4.44 &  $-$1.73 &&       &&  49.0 &&       &&       \nl
Fe I & 5398.29*&  4.44 &  $-$0.72 &&       && 100.7 &&       &&       \nl
Fe I & 5406.78 &  4.37 &  $-$1.40 &&       &&  65.7 &&       &&       \nl
Fe I & 5412.79 &  4.43 &  $-$1.80 &&       &&  57.6 &&       &&       \nl
Fe I & 5417.04 &  4.41 &  $-$1.42 &&       &&  67.0 &&       &&       \nl
Fe I & 5436.30 &  4.39 &  $-$1.36 &&  77.8 &&       &&       &&       \nl
Fe I & 5464.29 &  4.14 &  $-$1.62 &&  60.4 &&  75.0 &&       &&       \nl
Fe I & 5470.09*&  4.44 &  $-$1.60 &&  57.0 &&  55.7 &&       &&       \nl
Fe I & 5491.84 &  4.19 &  $-$2.24 &&  48.1 &&  36.5 &&  39.0 &&       \nl
Fe I & 5494.47 &  4.07 &  $-$1.96 &&  61.1 &&  68.9 &&  63.6 &&       \nl
Fe I & 5522.45*&  4.21 &  $-$1.47 &&  89.3 &&  86.9 &&  91.0 &&       \nl
Fe I & 5560.22*&  4.43 &  $-$1.10 &&  71.9 &&  80.1 &&       &&       \nl
Fe I & 5577.03*&  5.03 &  $-$1.49 &&       &&  33.1 &&       &&       \nl
Fe I & 5587.58 &  4.14 &  $-$1.70 &&  72.6 &&       &&  84.3 &&       \nl
Fe I & 5618.64 &  4.21 &  $-$1.34 &&       &&  90.2 && 100.0 &&  66.0 \nl
Fe I & 5619.61 &  4.39 &  $-$1.49 &&  68.6 &&       &&       &&  59.5 \nl
Fe I & 5650.00 &  5.10 &  $-$0.80 &&       &&  69.0 &&  65.3 &&       \nl
Fe I & 5651.48*&  4.47 &  $-$1.79 &&       &&  39.2 &&       &&       \nl
Fe I & 5661.35 &  4.28 &  $-$1.83 &&       &&  64.9 &&  55.4 &&       \nl
Fe I & 5701.56 &  2.56 &  $-$2.16 && 156.5 && 151.0 && 163.1 &&       \nl
Fe I & 5717.84 &  4.28 &  $-$0.98 &&       &&       &&       &&  89.3 \nl
Fe I & 5731.77*&  4.26 &  $-$1.10 && 101.6 &&  86.7 && 103.1 &&       \nl
Fe I & 5738.24 &  4.22 &  $-$2.24 &&  50.8 &&  40.9 &&  50.9 &&       \nl
Fe I & 5741.86*&  4.26 &  $-$1.69 &&       &&  61.5 &&  73.9 &&  71.2 \nl
Fe I & 5752.04*&  4.55 &  $-$0.92 &&  73.2 &&  82.3 &&  86.0 &&  87.3 \nl
Fe I & 5760.36 &  3.64 &  $-$2.46 &&  62.8 &&  67.3 &&  77.0 &&       \nl
Fe I & 5775.09 &  4.22 &  $-$1.11 &&  95.6 && 105.9 &&  97.2 &&       \nl
Fe I & 5778.46 &  2.59 &  $-$3.44 &&  82.2 &&  76.0 &&  76.8 &&       \nl
Fe I & 5784.67*&  3.40 &  $-$2.53 &&  77.0 &&  73.2 &&       &&       \nl
Fe I & 5793.92*&  4.22 &  $-$1.62 &&       &&  78.5 &&  77.2 &&       \nl
Fe I & 5835.11 &  4.26 &  $-$2.18 &&       &&       &&  58.1 &&       \nl
Fe I & 5852.23 &  4.55 &  $-$1.36 &&       &&       &&       &&  75.9 \nl
Fe I & 5855.09*&  4.61 &  $-$1.56 &&  54.1 &&  47.9 &&       &&  33.6 \nl
Fe I & 5856.10*&  4.29 &  $-$1.57 &&       &&  75.2 &&       &&  61.9 \nl
Fe I & 5858.78 &  4.22 &  $-$2.19 &&  36.8 &&  49.0 &&  41.9 &&  46.3 \nl
Fe I & 5859.60 &  4.55 &  $-$0.63 &&  97.5 && 105.5 && 118.4 &&  89.5 \nl
Fe I & 5862.37*&  4.55 &  $-$0.42 && 110.9 && 108.5 && 128.3 && 102.6 \nl
Fe I & 5881.28 &  4.61 &  $-$1.76 &&       &&  41.3 &&       &&       \nl
Fe I & 5905.68 &  4.65 &  $-$0.78 &&       &&       &&       &&  71.2 \nl
Fe I & 5927.80 &  4.65 &  $-$1.07 &&  66.5 &&  79.9 &&  69.9 &&  54.7 \nl
Fe I & 5929.68*&  4.55 &  $-$1.16 &&  83.2 &&  75.3 &&  71.9 &&       \nl
Fe I & 5930.19*&  4.65 &  $-$0.34 && 121.8 && 123.3 && 126.2 && 110.1 \nl
Fe I & 5934.66 &  3.93 &  $-$1.08 && 121.0 && 117.1 && 123.0 && 123.8 \nl
Fe I & 5956.71*&  0.86 &  $-$4.56 && 129.1 && 137.2 &&       && 128.4 \nl
Fe I & 5976.79*&  3.94 &  $-$1.30 && 111.1 && 109.5 && 109.1 &&       \nl
Fe I & 6003.02 &  3.88 &  $-$1.02 &&       &&       &&       && 123.1 \nl
Fe I & 6027.06*&  4.07 &  $-$1.20 &&  99.5 && 103.7 && 119.5 && 117.3 \nl
Fe I & 6056.01 &  4.73 &  $-$0.46 &&  88.0 && 104.5 && 102.8 && 110.9 \nl
Fe I & 6065.49*&  2.61 &  $-$1.49 && 189.6 && 192.7 && 199.9 &&       \nl
Fe I & 6078.50*&  4.79 &  $-$0.38 && 100.9 && 106.9 && 111.3 &&       \nl
Fe I & 6079.02*&  4.65 &  $-$0.97 &&  75.9 &&  80.9 &&  95.0 &&       \nl
Fe I & 6082.72 &  2.22 &  $-$3.53 && 108.1 &&       && 104.9 &&       \nl
Fe I & 6089.57 &  5.02 &  $-$0.87 &&  69.7 &&  70.5 &&  78.0 &&       \nl
Fe I & 6093.65*&  4.61 &  $-$1.32 &&  64.5 &&  73.5 &&       &&       \nl
Fe I & 6094.38 &  4.65 &  $-$1.56 &&       &&       &&  53.4 &&       \nl
Fe I & 6096.67*&  3.98 &  $-$1.76 &&  89.6 &&  78.1 &&  87.3 &&       \nl
Fe I & 6098.25 &  4.56 &  $-$1.81 &&       &&  42.5 &&  50.0 &&       \nl
Fe I & 6137.00 &  2.20 &  $-$2.91 && 147.6 &&       && 155.7 &&       \nl
Fe I & 6151.62*&  2.18 &  $-$3.26 && 109.3 && 114.5 && 115.5 && 115.5 \nl
Fe I & 6157.73 &  4.07 &  $-$1.26 &&       &&       &&       && 120.6 \nl
Fe I & 6165.36 &  4.14 &  $-$1.48 &&  85.0 &&  86.9 &&  90.6 &&  63.4 \nl
Fe I & 6173.34*&  2.22 &  $-$2.84 && 138.1 && 148.6 && 149.0 && 139.7 \nl
Fe I & 6187.99 &  3.94 &  $-$1.60 &&  99.1 &&  89.4 &&  93.8 &&       \nl
Fe I & 6200.32*&  2.61 &  $-$2.39 && 144.9 && 146.6 && 158.4 &&       \nl
Fe I & 6219.29 &  2.20 &  $-$2.39 &&       &&       &&       && 169.4 \nl
Fe I & 6232.65 &  3.65 &  $-$1.21 && 115.2 && 140.2 && 133.5 &&       \nl
Fe I & 6240.65 &  2.22 &  $-$3.23 && 109.7 &&       && 116.1 &&       \nl
Fe I & 6246.33*&  3.60 &  $-$0.73 && 146.7 && 158.2 && 156.0 && 145.9 \nl
Fe I & 6252.56*&  2.40 &  $-$1.64 &&       &&       && 206.1 &&       \nl
Fe I & 6265.14 &  2.18 &  $-$2.51 &&       &&       &&       && 187.6 \nl
Fe I & 6270.23 &  2.86 &  $-$2.55 && 115.4 && 102.8 && 116.5 && 102.2 \nl
Fe I & 6297.80*&  2.22 &  $-$2.70 && 157.9 &&       && 156.1 &&       \nl
Fe I & 6301.51*&  3.65 &  $-$0.72 && 161.0 && 162.2 && 178.5 &&       \nl
Fe I & 6322.69 &  2.59 &  $-$2.38 &&       &&       &&       && 135.4 \nl
Fe I & 6330.85 &  4.73 &  $-$1.22 &&       &&       &&       &&  65.5 \nl
Fe I & 6335.34 &  2.20 &  $-$2.28 &&       &&       &&       && 175.4 \nl
Fe I & 6380.75 &  4.19 &  $-$1.34 && 100.4 &&  98.5 && 104.5 && 110.6 \nl
Fe I & 6392.54*&  2.28 &  $-$3.97 &&  69.0 &&       &&  92.4 &&       \nl
Fe I & 6400.32 &  3.60 &  $-$0.23 && 211.9 && 226.6 && 241.2 &&       \nl
Fe I & 6411.66 &  3.65 &  $-$0.60 && 155.9 && 166.3 && 171.8 &&       \nl
Fe I & 6421.36 &  2.28 &  $-$1.98 && 194.2 && 206.6 && 229.0 &&       \nl
Fe I & 6481.88 &  2.28 &  $-$2.94 && 143.9 && 142.7 &&       && 123.1 \nl
Fe I & 6498.95 &  0.96 &  $-$4.66 && 133.7 && 136.1 && 151.2 && 155.9 \nl
Fe I & 6518.37*&  2.83 &  $-$2.56 && 111.7 && 109.6 && 112.3 &&       \nl
Fe I & 6533.94 &  4.56 &  $-$1.28 &&  80.8 &&       &&  73.6 &&       \nl
Fe I & 6574.25*&  0.99 &  $-$4.96 &&       &&       &&       && 104.3 \nl
Fe I & 6581.22 &  1.48 &  $-$4.68 &&       &&       && 119.3 && 103.8 \nl
Fe I & 6593.88*&  2.43 &  $-$2.30 && 164.9 && 164.0 && 171.0 && 168.0 \nl
Fe I & 6608.04 &  2.28 &  $-$3.96 &&  67.2 &&  76.2 &&  75.9 &&  66.3 \nl
Fe I & 6625.04 &  1.01 &  $-$5.32 &&       &&       && 117.1 && 123.7 \nl
Fe I & 6627.56 &  4.55 &  $-$1.50 &&       &&  58.5 &&  62.9 &&       \nl
Fe I & 6633.76*&  4.56 &  $-$0.81 &&  94.3 &&       &&       &&       \nl
Fe I & 6703.58 &  2.76 &  $-$3.00 &&  84.1 &&  98.1 &&       &&  88.1 \nl
Fe I & 6713.74 &  4.79 &  $-$1.41 &&  58.7 &&       &&  49.2 &&       \nl
Fe I & 6725.36*&  4.10 &  $-$2.21 &&  41.9 &&       &&       &&  46.6 \nl
Fe I & 6726.67*&  4.61 &  $-$1.05 &&  82.5 &&  77.6 &&  88.2 &&  74.5 \nl
Fe I & 6733.15 &  4.64 &  $-$1.44 &&  55.5 &&  53.8 &&  50.6 &&  67.4 \nl
Fe I & 6750.16*&  2.42 &  $-$2.58 && 144.7 && 152.3 && 162.0 && 146.5 \nl
Fe I & 6786.86*&  4.19 &  $-$1.90 &&  64.6 &&  55.9 &&  58.6 &&       \nl
Fe I & 6806.86 &  2.73 &  $-$3.14 &&       &&       &&       && 109.0 \nl
Fe I & 6810.27 &  4.61 &  $-$1.00 &&       &&       &&       && 100.6 \nl
Fe I & 6820.37 &  4.64 &  $-$1.16 &&       &&       &&       &&  81.9 \nl
Fe I & 6837.01 &  4.59 &  $-$1.71 &&  42.9 &&  58.1 &&  46.2 &&       \nl
Fe I & 6839.84 &  2.56 &  $-$3.35 &&  95.6 &&  97.0 &&  98.3 && 106.1 \nl
Fe I & 6843.66 &  4.55 &  $-$0.86 &&  85.0 &&       && 100.9 &&  92.7 \nl
Fe I & 6858.16 &  4.61 &  $-$0.95 &&       &&  79.1 &&       &&  97.7 \nl
Fe I & 6861.95 &  2.42 &  $-$3.78 &&  67.0 &&  77.6 &&  84.4 &&       \nl
Fe I & 6898.29 &  4.22 &  $-$2.08 &&  44.4 &&  43.8 &&  64.2 &&       \nl
Fe I & 6916.69 &  4.15 &  $-$1.35 &&  82.3 &&  99.5 && 113.5 &&       \nl
Fe I & 6945.20 &  2.42 &  $-$2.46 &&       &&       &&       && 175.0 \nl
Fe I & 6951.25 &  4.56 &  $-$1.05 &&       &&       &&       &&  99.8 \nl
Fe I & 6971.94*&  3.02 &  $-$3.34 &&  63.4 &&  60.7 &&  64.1 &&       \nl
Fe I & 6978.86 &  2.48 &  $-$2.49 &&       && 142.3 && 143.0 &&       \nl
Fe I & 6988.53*&  2.40 &  $-$3.42 && 108.2 && 103.4 && 108.4 && 109.3 \nl
Fe I & 7007.97 &  4.18 &  $-$1.80 &&  65.0 &&  60.7 &&  64.7 &&  82.0 \nl
Fe I & 7010.35 &  4.59 &  $-$1.86 &&       &&  46.7 &&       &&  52.1 \nl
Fe I & 7022.96*&  4.19 &  $-$1.11 && 110.9 && 100.2 && 108.6 && 106.2 \nl
Fe I & 7024.07 &  4.07 &  $-$1.94 &&       &&  78.3 &&       &&       \nl
Fe I & 7090.39 &  4.23 &  $-$1.06 &&       &&       &&       && 133.6 \nl
Fe I & 7114.56*&  2.69 &  $-$3.93 &&  46.5 &&       &&       &&       \nl
Fe I & 7118.10*&  5.01 &  $-$1.52 &&       &&  35.4 &&       &&       \nl
Fe I & 7132.99 &  4.07 &  $-$1.66 &&  81.9 &&  87.2 &&  89.3 &&       \nl
Fe I & 7142.52 &  4.95 &  $-$0.93 &&       &&       &&  81.9 &&  89.1 \nl
Fe I & 7151.47 &  2.48 &  $-$3.58 && 101.2 &&       && 107.0 && 101.3 \nl
Fe I & 7189.16*&  3.07 &  $-$2.77 &&       &&  90.7 &&       &&       \nl
Fe I & 7284.84 &  4.14 &  $-$1.63 &&  89.7 &&       &&  96.1 &&       \nl
Fe I & 7306.57 &  4.18 &  $-$1.55 &&  78.8 &&  82.5 &&  93.0 &&       \nl
Fe I & 7401.69 &  4.19 &  $-$1.60 &&  71.2 &&       &&  93.5 && 100.2 \nl
Fe I & 7411.16 &  4.28 &  $-$0.48 && 153.3 && 153.6 &&       &&       \nl
Fe I & 7418.67 &  4.14 &  $-$1.44 &&  78.5 &&  85.3 &&  91.1 && 103.0 \nl
Fe I & 7421.56*&  4.64 &  $-$1.69 &&  37.3 &&       &&       &&  36.0 \nl
Fe I & 7430.54*&  2.59 &  $-$3.82 &&  80.5 &&  79.4 &&  91.1 &&       \nl
Fe I & 7447.40 &  4.95 &  $-$0.95 &&       &&  60.7 &&       &&  63.0 \nl
Fe I & 7461.53 &  2.56 &  $-$3.45 &&       &&  85.0 &&       && 100.2 \nl
Fe I & 7491.66 &  4.30 &  $-$1.01 &&  99.6 &&       && 111.2 &&       \nl
Fe I & 7540.44 &  2.73 &  $-$3.75 &&       &&       &&       &&  74.2 \nl
Fe I & 7547.90 &  5.10 &  $-$1.11 &&       &&       &&       &&  44.1 \nl
Fe I & 7568.91*&  4.28 &  $-$0.90 && 121.2 &&       && 133.2 && 125.7 \nl
Fe I & 7582.12*&  4.95 &  $-$1.60 &&       &&  30.0 &&       &&       \nl
Fe I & 7583.80 &  3.02 &  $-$1.93 && 150.8 && 149.4 && 165.5 && 171.2 \nl
Fe I & 7719.05*&  5.03 &  $-$0.96 &&       &&  55.0 &&       &&  64.9 \nl
Fe I & 7723.21 &  2.28 &  $-$3.62 && 100.3 &&       && 108.2 && 111.9 \nl
Fe I & 7745.52 &  5.08 &  $-$1.14 &&       &&       &&       &&  65.0 \nl
Fe I & 7751.11*&  4.99 &  $-$0.74 &&  82.6 &&  85.6 &&  94.8 &&       \nl
Fe I & 7807.91 &  4.99 &  $-$0.51 &&  84.5 &&  93.1 &&       &&       \nl
Fe II& 5414.08 &  3.22 &  $-$3.61 &&       &&  42.4 &&       &&       \nl
Fe II& 5425.26 &  3.20 &  $-$3.27 &&  55.4 &&  61.0 &&       &&       \nl
Fe II& 5991.38 &  3.15 &  $-$3.55 &&       &&  62.9 &&  65.1 &&       \nl
Fe II& 6084.10 &  3.20 &  $-$3.80 &&       &&       &&  52.9 &&       \nl
Fe II& 6149.25 &  3.89 &  $-$2.72 &&  52.6 &&  50.0 &&  51.7 &&  52.0 \nl
Fe II& 6247.56 &  3.87 &  $-$2.32 &&  61.5 &&  65.0 &&  69.1 &&  74.3 \nl
Fe II& 6369.46 &  2.89 &  $-$4.21 &&  27.2 &&  30.2 &&  36.5 &&  40.2 \nl
Fe II& 6416.93 &  3.89 &  $-$2.70 &&  52.4 &&  60.1 &&       &&       \nl
Fe II& 6432.68 &  2.89 &  $-$3.58 &&       &&       &&       &&  67.7 \nl
Fe II& 6516.08 &  2.89 &  $-$3.38 &&       &&       &&  98.9 &&  77.2 \nl
Ni I & 5468.11 &  3.85 &  $-$1.63 &&  28.8 &&       &&  31.0 &&       \nl
Ni I & 5578.73 &  1.68 &  $-$2.57 && 122.5 && 130.1 && 137.1 &&       \nl
Ni I & 5587.87 &  1.93 &  $-$2.39 && 116.4 && 125.9 && 124.5 &&       \nl
Ni I & 5593.75 &  3.90 &  $-$0.78 &&  66.0 &&  66.4 &&       &&       \nl
Ni I & 5643.09 &  4.16 &  $-$1.25 &&       &&  33.7 &&       &&       \nl
Ni I & 5748.36 &  1.68 &  $-$3.25 &&  98.6 && 101.4 && 102.3 &&       \nl
Ni I & 5760.84 &  4.10 &  $-$0.81 &&  82.8 &&  77.6 &&  73.1 &&       \nl
Ni I & 5805.23 &  4.17 &  $-$0.60 &&  82.3 &&       &&       &&       \nl
Ni I & 5847.01 &  1.68 &  $-$3.44 &&  92.6 &&  94.1 && 103.2 &&       \nl
Ni I & 5996.74 &  4.23 &  $-$1.06 &&  74.8 &&  47.8 &&       &&  41.7 \nl
Ni I & 6053.69 &  4.23 &  $-$1.07 &&  54.9 &&  75.6 &&  78.9 &&       \nl
Ni I & 6086.29 &  4.26 &  $-$0.47 &&  70.4 &&  73.4 &&  89.4 &&       \nl
Ni I & 6108.13 &  1.68 &  $-$2.47 &&       &&       &&       && 151.1 \nl
Ni I & 6111.08 &  4.09 &  $-$0.83 &&       &&       &&       &&  75.0 \nl
Ni I & 6128.98 &  1.68 &  $-$3.39 && 100.4 &&  97.1 &&  96.3 &&  74.2 \nl
Ni I & 6130.14 &  4.26 &  $-$0.98 &&       &&       &&  51.3 &&  26.6 \nl
Ni I & 6176.82 &  4.09 &  $-$0.24 && 100.2 && 100.7 && 106.6 &&       \nl
Ni I & 6177.25 &  1.83 &  $-$3.60 &&  70.6 &&       &&  69.5 &&       \nl
Ni I & 6186.72 &  4.10 &  $-$0.90 &&  67.4 &&  57.8 &&  65.7 &&       \nl
Ni I & 6204.61 &  4.09 &  $-$1.15 &&  60.6 &&  59.1 &&  63.6 &&       \nl
Ni I & 6327.60 &  1.68 &  $-$3.08 &&       &&       &&       && 122.7 \nl
Ni I & 6378.26 &  4.15 &  $-$0.82 &&  62.7 &&  67.5 &&  64.7 &&  55.6 \nl
Ni I & 6384.67 &  4.15 &  $-$1.00 &&  63.4 &&  72.1 &&  61.5 &&       \nl
Ni I & 6482.81 &  1.93 &  $-$2.78 && 105.0 && 108.7 && 125.4 && 112.6 \nl
Ni I & 6532.88 &  1.93 &  $-$3.42 &&  82.2 &&  88.2 &&  84.5 &&       \nl
Ni I & 6586.32 &  1.95 &  $-$2.78 && 111.7 && 111.6 && 110.0 && 104.8 \nl
Ni I & 6598.61 &  4.23 &  $-$0.93 &&  50.9 &&  45.9 &&  55.7 &&       \nl
Ni I & 6635.14 &  4.42 &  $-$0.75 &&  65.8 &&  60.0 &&  48.2 &&  77.2 \nl
Ni I & 6767.78 &  1.83 &  $-$2.06 && 140.7 && 153.3 && 161.9 &&       \nl
Ni I & 6772.32 &  3.66 &  $-$0.96 &&  84.0 &&  88.9 && 107.9 &&       \nl
Ni I & 6842.04 &  3.66 &  $-$1.44 &&  66.5 &&  71.3 &&  67.7 &&       \nl
Ni I & 7001.55 &  1.93 &  $-$3.65 &&  64.3 &&  61.7 &&  67.8 &&  49.7 \nl
Ni I & 7030.02 &  3.54 &  $-$1.70 &&  46.3 &&       &&       &&  50.7 \nl
Ni I & 7034.38 &  3.54 &  $-$2.10 &&  61.5 &&  48.6 &&  59.7 &&       \nl
Ni I & 7110.91 &  1.93 &  $-$2.91 && 132.5 && 130.5 && 122.3 && 116.6 \nl
Ni I & 7327.65 &  3.80 &  $-$1.75 &&  36.7 &&  38.2 &&       &&       \nl
Ni I & 7381.94 &  5.36 &  $-$0.05 &&       &&       &&       &&  30.3 \nl
Ni I & 7401.12 &  5.36 &  $-$0.18 &&  51.7 &&  46.5 &&       &&       \nl
Ni I & 7422.29 &  3.63 &  $-$0.29 && 146.9 && 160.6 && 157.9 && 162.9 \nl
Ni I & 7525.12 &  3.63 &  $-$0.67 &&       &&       &&       && 119.0 \nl
Ni I & 7555.61 &  3.85 &  $-$0.12 &&       &&       &&       && 134.8 \nl
Ni I & 7574.05 &  3.83 &  $-$0.61 && 108.1 && 118.3 && 114.9 && 131.4 \nl
Ni I & 7715.58 &  3.70 &  $-$0.98 &&       &&       &&       && 110.6 \nl
Ni I & 7727.62 &  3.68 &  $-$0.30 && 127.5 && 124.3 && 135.0 &&       \nl
Ni I & 7797.59 &  3.30 &  $-$0.82 && 125.1 &&       && 123.8 &&       \nl
Ni I & 7863.79 &  4.54 &  $-$0.94 &&       &&       &&       &&  54.2 \nl
Zn I & 6362.35 &  5.79 &     0.14 &&  67.3 &&  60.9 &&  77.1 &&       \nl
Y II & 5509.91 &  0.99 &  $-$1.01 && 110.7 && 121.6 &&       &&       \nl
Ba II& 5853.69 &  0.60 &  $-$1.00 && 127.2 && 127.5 && 129.7 && 126.1 \nl
Ba II& 6141.75 &  0.70 &     0.00 && 169.4 && 184.3 && 199.7 && 155.4 \nl
Ba II& 6496.91 &  0.60 &  $-$0.38 && 171.0 && 185.3 && 177.9 && 199.7 \nl
\enddata
\tablenotetext{a}{As described in the text, all \eqw are measured from 
original spectra, but for star 3046 they are measured on convolved 
spectra and converted to the system of the other \eqw.}
\tablenotetext{b}{Fe I lines marked with an asterisk are those adopted as very
clean lines; see text in Section 3.4}
\end{deluxetable}

%
%
\begin{deluxetable}{lrrr}
\tablenum{3}
\tablewidth{0pt}
\scriptsize
\tablecaption{Adopted Model Atmosphere Parameters}
\label{tab2}
\tablehead{\colhead{Star ID} & \colhead{\teff} & 
\colhead{log(g)} & \colhead{$v_t$} \nl
\colhead{} & \colhead{(K)} & \colhead{(dex)} &
\colhead{(\kms)}
}
\startdata
Red HB Stars \nl
5422   &   4580 &  2.3  &   1.25 \nl
3025   &   4650 &  2.3  &   1.32 \nl
3014   &   4630 &  2.0  &   1.45 \nl
3046   &   4580 &  2.0  &   1.34 \nl
\enddata
\end{deluxetable}

%
%
\begin{deluxetable}{lrrrrr}
\tablenum{4}
\tablewidth{0pt}
\tabcolsep 2pt
\scriptsize
\tablecaption{Sensitivity of Abundances To Atmospheric Parameters}
\label{tab2}
\tablehead{
\colhead{}& \colhead{$\Delta T_{\rm eff}$} & \colhead{$\Delta log g$} & 
\colhead{$\Delta [A/H]$} & \colhead{$\Delta v_t$} & \colhead{Total} \nl
\colhead{Change} & \colhead{+100~K} & \colhead{+0.2~dex} & \colhead{+0.1~dex} 
& \colhead{+0.2~km$\,$s$^{-1}$} & \colhead{}
}
\startdata
[O/Fe]    & $-$0.352  & $-$0.131 &  $-$0.142 &  $-$0.073  &   0.41 \nl
[Na/Fe]   &    0.000  & $-$0.058 &  $-$0.030 &  $-$0.003  &   0.06 \nl
[Mg/Fe]   &    0.050  & $-$0.078 &     0.037 &     0.047  &   0.11 \nl
[Al/Fe]   & $-$0.125  & $-$0.078 &  $-$0.063 &  $-$0.013  &   0.16 \nl
[Si/Fe]   & $-$0.100  &    0.019 &     0.008 &     0.045  &   0.11 \nl
[Ca/Fe]   &    0.062  & $-$0.040 &  $-$0.021 &  $-$0.035  &   0.08 \nl
[Sc/Fe]II &    0.123  & $-$0.031 &  $-$0.009 &  $-$0.008  &   0.13 \nl
[Ti/Fe]I  &    0.134  & $-$0.016 &  $-$0.018 &  $-$0.014  &   0.14 \nl
[V/Fe]    &    0.128  & $-$0.013 &  $-$0.028 &  $-$0.039  &   0.14 \nl
[Cr/Fe]   &    0.083  & $-$0.012 &  $-$0.013 &     0.029  &   0.09 \nl
[Mn/Fe]   &    0.037  & $-$0.028 &  $-$0.005 &  $-$0.033  &   0.06 \nl
[Fe/H]I   &    0.050  &    0.018 &     0.013 &  $-$0.087  &   0.10 \nl
[Fe/H]II  & $-$0.123  &    0.111 &     0.042 &  $-$0.062  &   0.18 \nl
[Ni/Fe]   & $-$0.027  &    0.034 &     0.008 &  $-$0.001  &   0.04 \nl
[Ba/Fe]II & $-$0.050  &    0.042 &     0.020 &  $-$0.093  &   0.12 \nl
\enddata
\end{deluxetable}

%
%
\begin{deluxetable}{llrllrllrllrllrllc}
\tablenum{5}
\tablewidth{0pt}
\tabcolsep 3pt 
\scriptsize
\tablecaption{Abundances for Four Red HB Stars in NGC 6528}
\label{tab4}
\tablehead{\colhead{\bf Ion} && 
\multicolumn{2}{c}{\bf Star 5422} &&
\multicolumn{2}{c}{\bf Star 3025} && 
\multicolumn{2}{c}{\bf Star 3014} &&
\multicolumn{2}{c}{\bf Star 3046} &&
\multicolumn{2}{c}{\bf $\zeta$ Cygni} && 
\colhead{\bf log n} \nl
\colhead{} && 
\colhead{\# of} & \colhead{abundance} && 
\colhead{\# of} & \colhead{abundance} && 
\colhead{\# of} & \colhead{abundance} && 
\colhead{\# of} & \colhead{abundance} && 
\colhead{\# of} & \colhead{abundance} && 
\colhead{(sun)} \nl
\colhead{} && 
\colhead{lines} & \colhead{(dex)} && 
\colhead{lines} & \colhead{(dex)} && 
\colhead{lines} & \colhead{(dex)} && 
\colhead{lines} & \colhead{(dex)} && 
\colhead{lines} & \colhead{(dex)} && 
\colhead{} 
}
\startdata
[Fe/H] \nl
Fe I  &&93 & +0.05 (0.16) &&100&   +0.08 (0.11) &&90 &   +0.09 (0.14) &&70 &   
+0.04 (0.24) && 27 & +0.05\tablenotemark{b} (0.20) &&7.50 \nl
Fe II && 5 & +0.12 (0.14) && 7 &   +0.15 (0.16) && 6 &   +0.15 (0.21) &&5  &   
+0.10 (0.04) && 1 & $-$0.06 &&7.44 \nl
 & \nl
[El/Fe] \nl
O I\tablenotemark{a}   &&4  & +0.18  (0.29) &&4  & +0.15 (0.12) &&3  &$-$0.08 
(0.05)&&   &              &&   &              && 8.62 \nl
O I\tablenotemark{a} (nLTE)&& 4&+0.16 (0.27)&&4  & +0.13 (0.11) &&3  &$-$0.09 
(0.03)&&   &              && 3 & +0.38 (0.18) && 8.62 \nl
Na I                   &&4  & +0.44  (0.13) &&4  & +0.51 (0.16) &&4  & +0.54 
(0.16) &&2  & +0.19 (0.01) &&   &              && 6.33 \nl
Na I (nLTE)            &&4  & +0.42  (0.12) &&4  & +0.47 (0.16) &&4  & +0.51 
(0.15) &&2  & +0.20 (0.04) &&   &              && 6.33 \nl
Mg I	               &&3  & +0.11  (0.17) &&3  & +0.05 (0.20)&&3  & +0.27 
(0.25) &&   &              && 1 & +0.20        && 7.54 \nl
Si I                   &&14 & +0.44  (0.25) &&14 & +0.39 (0.20) &&13 & +0.37 
(0.22) &&7  & +0.24 (0.12) && 2 & +0.11 (0.26) && 7.54 \nl
Ca I	               &&13 & +0.23  (0.26) &&11 & +0.27 (0.14) &&12 & +0.28 
(0.18) &&8  & +0.14 (0.17) && 3 & +0.07 (0.04) && 6.16 \nl
Sc II\tablenotemark{a} &&5  &$-$0.07 (0.06) &&6  &$-$0.09 (0.08) &&5  & +0.09 
(0.06) &&3  &$-$0.13 (0.31)&& 1 & +0.12        && 2.96 \nl
Ti I	               &&10 & +0.08  (0.29) &&9  & +0.08 (0.10) &&10 & +0.05 
(0.25) &&6  &$-$0.08 (0.30)&& 1 & $-$0.28      && 4.98 \nl
V I                    &&6  &$-$0.20 (0.13) &&6  &$-$0.25 (0.14) &&6  &$-$0.08 
(0.21) &&6  &$-$0.29 (0.25)&&   &              && 4.00 \nl
Cr I	               &&9  & +0.05  (0.25) &&13 & +0.01 (0.22) &&11 &$-$0.06 
(0.23)&&4  & +0.00 (0.39) && 1 & +0.02        && 5.62 \nl
Mn I                   &&1  &$-$0.39          &&4  &$-$0.42 (0.18) 
&&   &  &&3  &$-$0.29 (0.14) &&   &              && 5.39 \nl
Ni I	               &&36 & +0.15  (0.21) &&32 & +0.14 (0.20) &&31 & +0.05 
(0.17) &&20 & +0.06 (0.29) && 9 &$-$0.08 (0.17)&& 6.25 \nl
Ba II                  &&3  & +0.18  (0.20) &&3  & +0.24 (0.14) &&3  & +0.08  
(0.01)&&3  & +0.07 (0.38) &&   &              && 2.21 \nl
\enddata
\tablenotetext{a}{O I and Sc II are calculated with respect to Fe II, all other 
ions are with
respect to Fe I.}
\tablenotetext{b}{Abundance for $\zeta$ Cyg are those derived from the convolved
spectra (see paper I).}
\end{deluxetable}

%
%

\begin{deluxetable}{lclclclrr}
\tablenum{6}
\tablewidth{0pt}
\scriptsize
\tablecaption{Mean Abundances for NGC 6528 and Comparison With Results For 
NGC 6553 and For Baade's Window} 
\label{tab6}
\tablehead{
Ion & \colhead{NGC 6528} & \colhead{$\sigma$} & \colhead{NGC 6553} 
 & \colhead{$\sigma$} &\colhead{NGC 6553} & \colhead{$\sigma$} 
 & \colhead{Mean BW\tablenotemark{a}} &\colhead{$\sigma$} \nl    
 & &  & \colhead{Paper I}  &                    &\colhead{Barbuy et al.} 
&               &                                    & \nl
 & \colhead{(dex)}    & \colhead{(dex)}    & \colhead{(dex)}    
& \colhead{(dex)}    & \colhead{(dex)}    & \colhead{(dex)}    
& \colhead{(dex)} & \colhead{(dex)} \nl 
}
\startdata
[Fe/H] \nl
Fe I  & +0.07   & 0.02 &$-$0.16 & 0.08 & $-$0.55 & 0.20 & $-$0.33 & \nl
Fe II & +0.13   & 0.02 &$-$0.18 & 0.10 &         &      &         & \nl
 & \nl
[El/Fe]& \nl
O I   &   +0.07 & 0.11 &   +0.50 & 0.13 &         &      &   +0.03 & 0.18 \nl
Na I  &   +0.40 & 0.12 &         &      &   +0.65 & 0.05 &   +0.21 & 0.37 \nl
Mg I  &   +0.14 & 0.09 & (+0.41)\tablenotemark{b} & (0.10) &   +0.33 & 0.13 &   
+0.35 & 0.14 \nl
Si I  &   +0.36 & 0.07 &   +0.14 & 0.18 &   +0.35 & 0.05 &   +0.18 & 0.24 \nl
Ca I  &   +0.23 & 0.06 &   +0.26 & 0.09 &   +0.32 & 0.12 &   +0.14 & 0.17 \nl
Sc II & $-$0.05 & 0.10 & ($-$0.12) & (0.18) &         &      &+0.29 & 0.20 \nl
Ti I  &   +0.03 & 0.07 &   +0.19 & 0.06 &   +0.51 & 0.09 &   +0.34 & 0.10 \nl
V I   & $-$0.20 & 0.09 &         &      &         &      &   +0.06 & 0.19 \nl
Cr I  &    0.00 & 0.04 &   +0.04 & 0.09 &         &      & $-$0.04 & 0.19 \nl
Mn I  & $-$0.37 & 0.07 &         &      &         &      &         &      \nl
Ni I  &   +0.10 & 0.05 &   +0.01 & 0.07 &         &      & $-$0.04 & 0.08 \nl
Ba II &   +0.14 & 0.07 &         &      & $-$0.10 & 0.30 &   +0.20 & 0.28 \nl
\enddata
\tablenotetext{a}{Abundances for Baade's Window are from the 11 giants
studied by McWilliam \& Rich (1994).}
\tablenotetext{b}{Parentheses indicate ions where a maximum of one line
has been observed per star.}
\end{deluxetable}

\clearpage

\begin{figure}
\epsscale{1.0}
\plotone{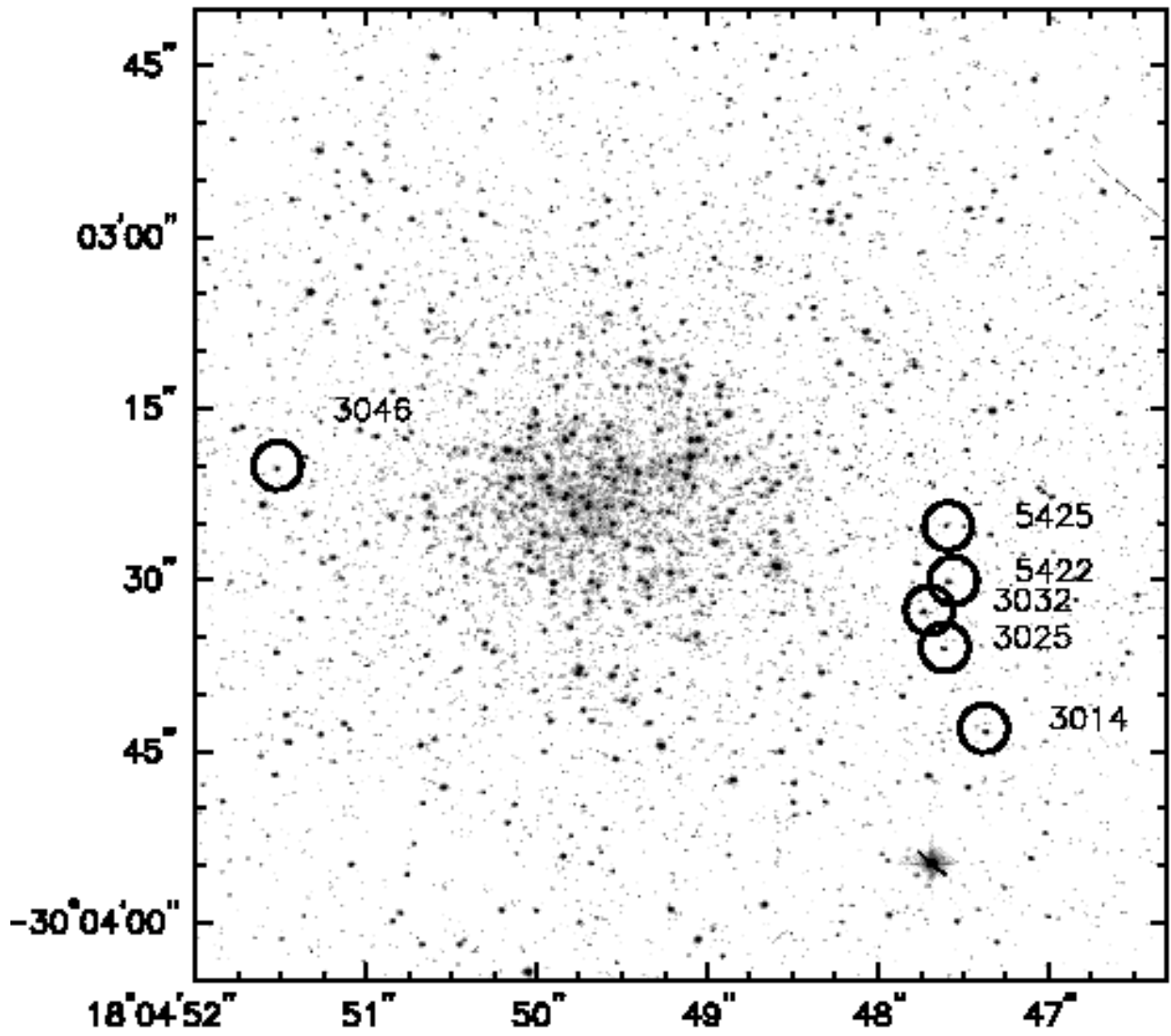}
\caption[carretta.fig1.ps]{The stars in NGC~6528 studied here are marked on 
this subset from a 100 sec WFPC2 image from the HST Archive taken with the 
F555W filter.  
\label{fig1}}
\end{figure}

\begin{figure}
\epsscale{1.0}
\plotone{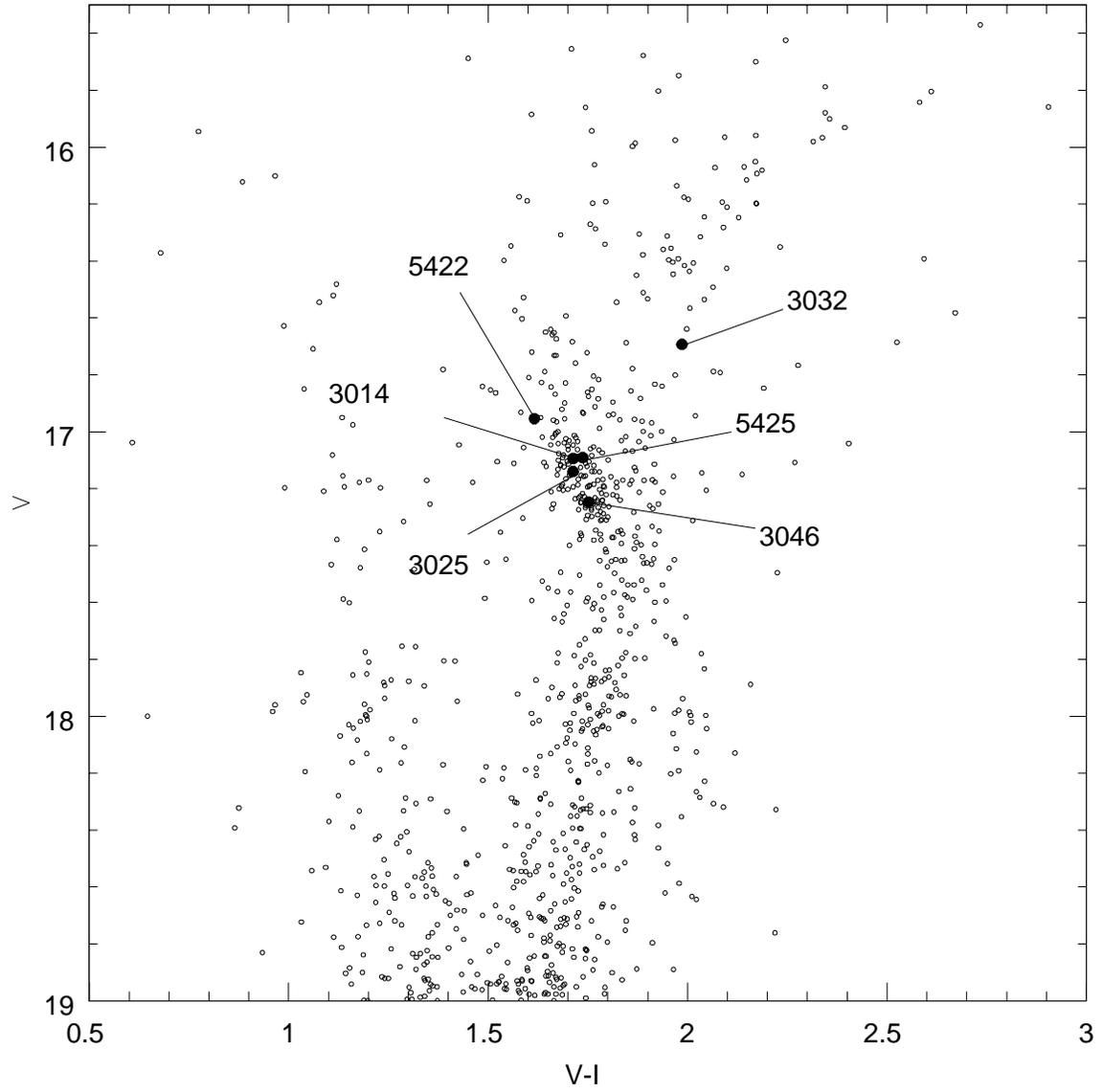}
\caption[carretta.fig2.ps]{$V,V-I$ color-magnitude diagram of NGC 6528. Stars studied in
the present work are indicated with larger, filled dot and labelled in Figure.
Note that stars 3032 and 5425 were not used in the abundance analysis, due to
the low S/N of their spectra.
\label{fig2}}
\end{figure}

\begin{figure}
\epsscale{1.0}
\plotone{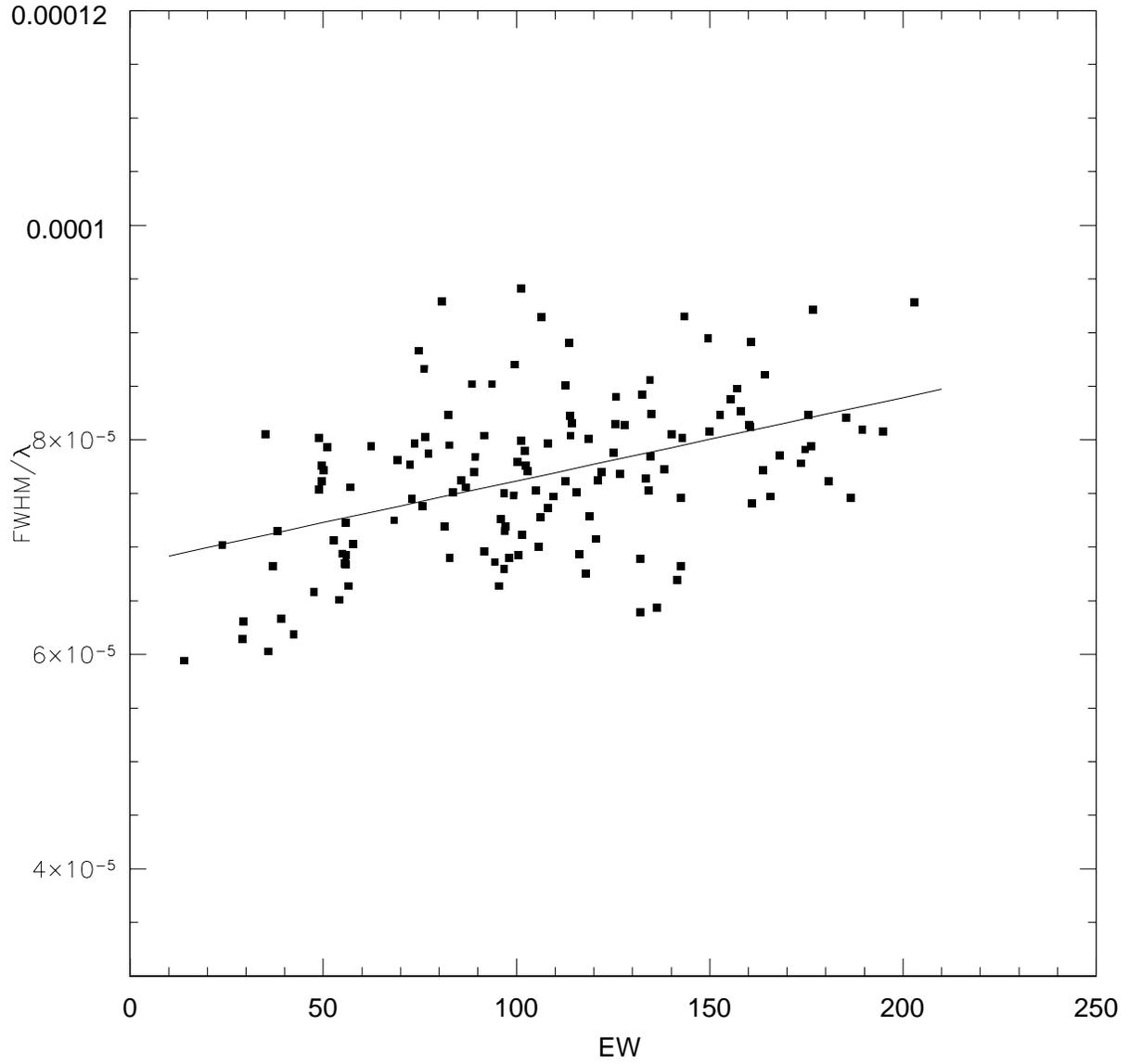}
\caption[carretta.fig3.ps]{Fiducial relation between the FWHM of lines and their
EWs for star 3046, observed at lower resolution.
\label{fig3}}
\end{figure}

\begin{figure}
\epsscale{1.0}
\plotone{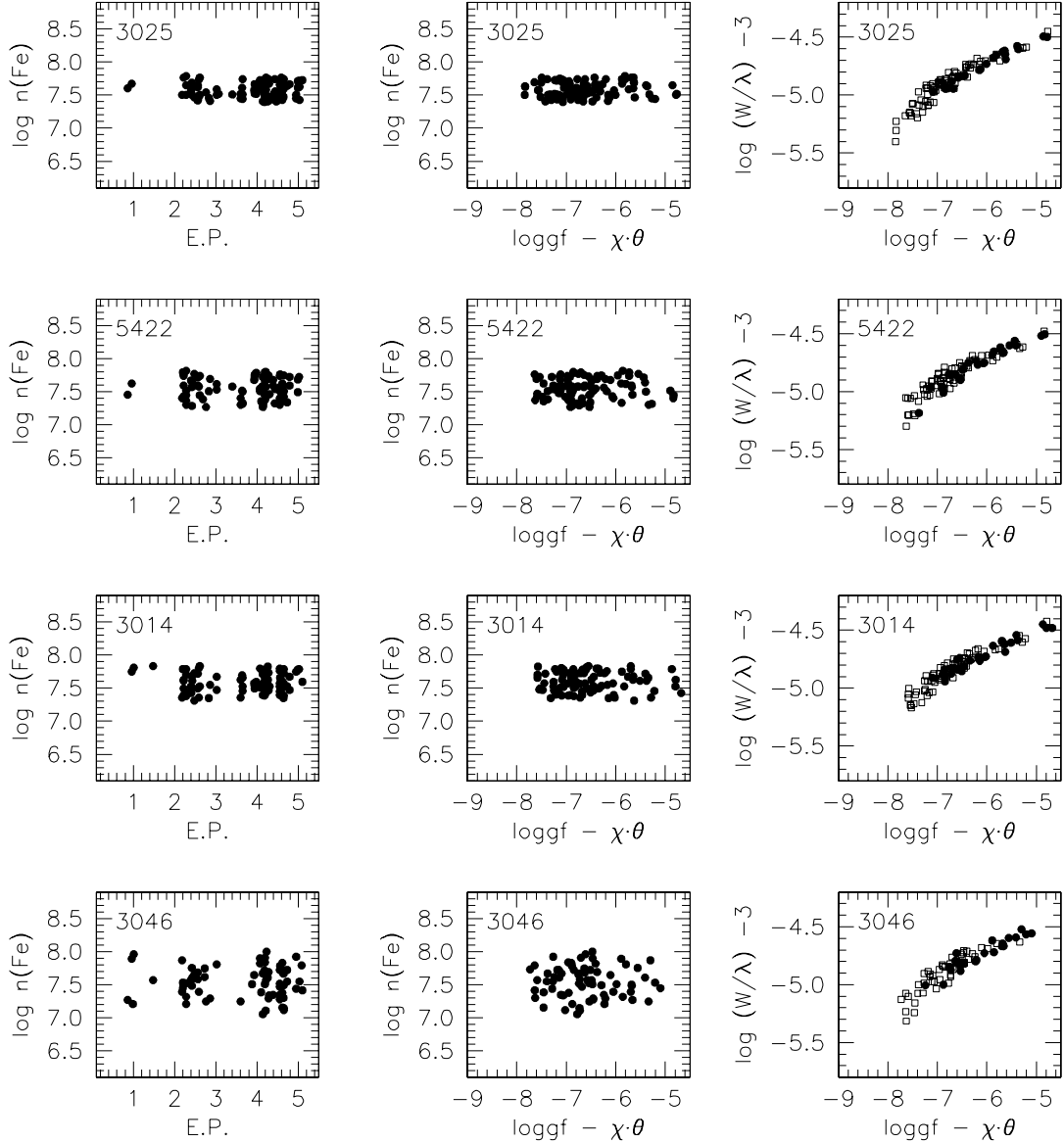}
\caption[carretta.fig4.ps]{Abundances deduced from neutral Fe~I lines as a function of
excitation potential of the lower level of the transition (panels on the left) and
of the equivalent widths (central panels) for the NGC 6528 RHB stars.
In the panels on the right, curves-of-growth are also displayed, with open squares
for lines with $\chi> 3$ eV and filled circles for lines having $\chi\leq 3$ eV.
\label{fig4}}
\end{figure}

\begin{figure}
\epsscale{1.0}
\plotone{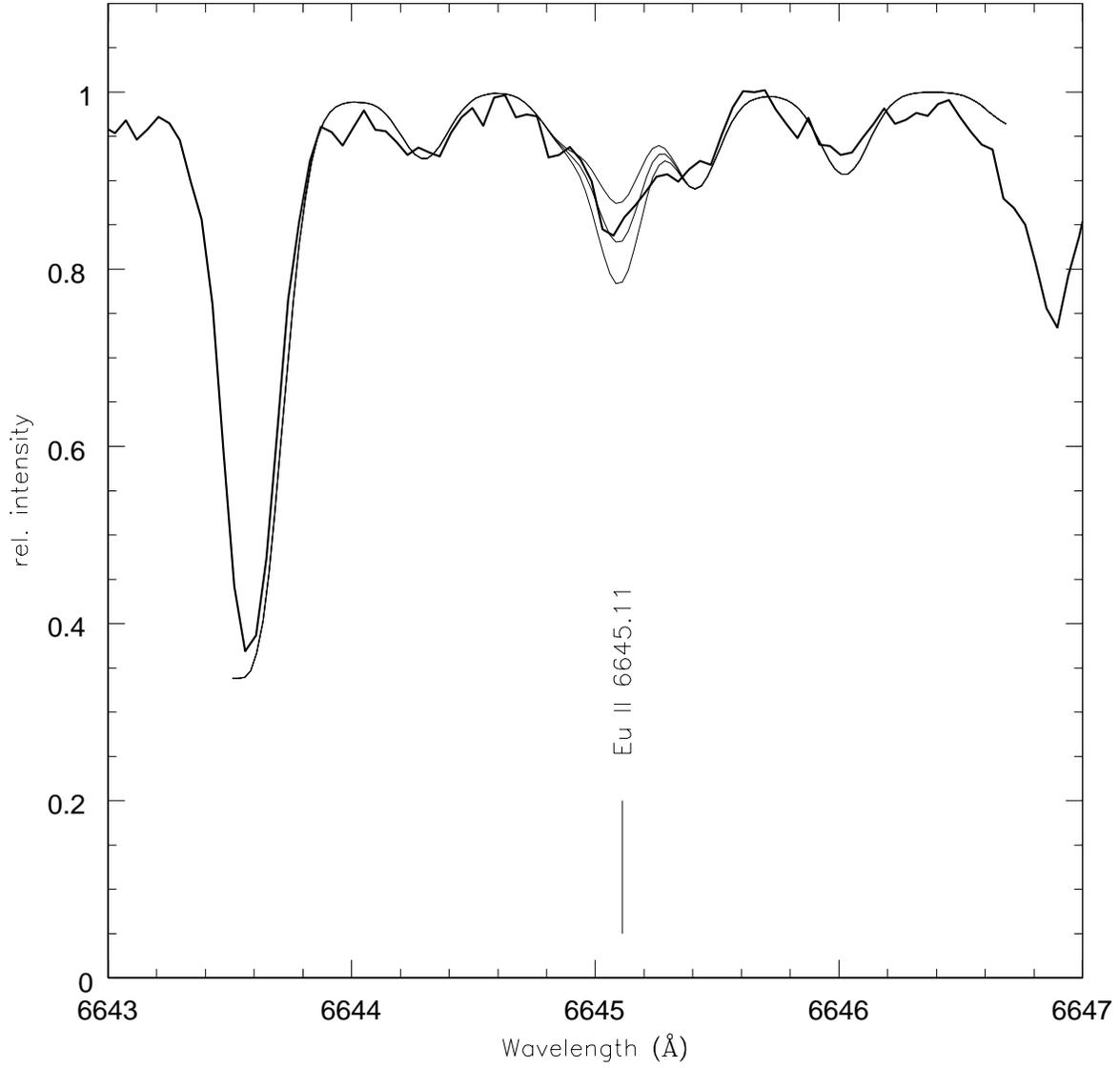}
\caption[carretta.fig5.ps]{Comparison between the observed spectrum in the
region of the Eu II line at 6645.11~\AA (average of all stars in NGC 6528: thick
line) and synthetic spectra computed for [Eu/Fe]= 0.0. 0.25, 0.50 (thin lines).
From this comparison, evidence for a mild ([Eu/Fe] $\simeq 0.2$) overabundance
of Eu is found.
\label{fig5}}
\end{figure}

\begin{figure}
\epsscale{1.0}
\plotone{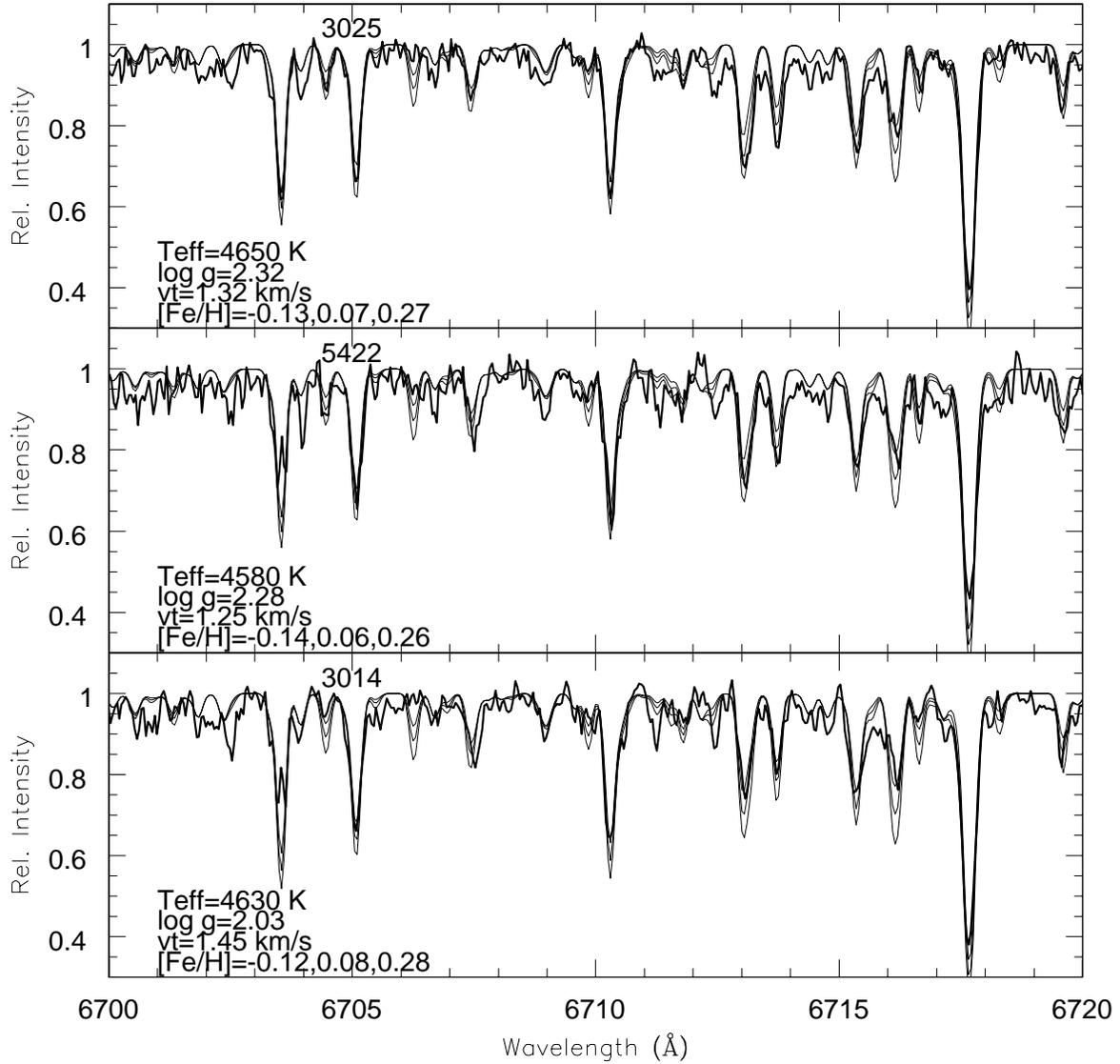}
\caption[carretta.fig6.ps]{Spectral synthesis of the region around the Li doublet for
the 3 RHB stars of NGC 6528 with spectra taken in the run of June 2000.
Thick lines are the observed spectra, while the thin lines are predictions for model
atmospheres, with the parameters indicated in each box, for abundances of 
[Fe/H] listed at the bottom of each panel. 
Wavelengths have been shifted into the rest frame.
\label{fig6}}
\end{figure}

\begin{figure}
\epsscale{1.0}
\plotone{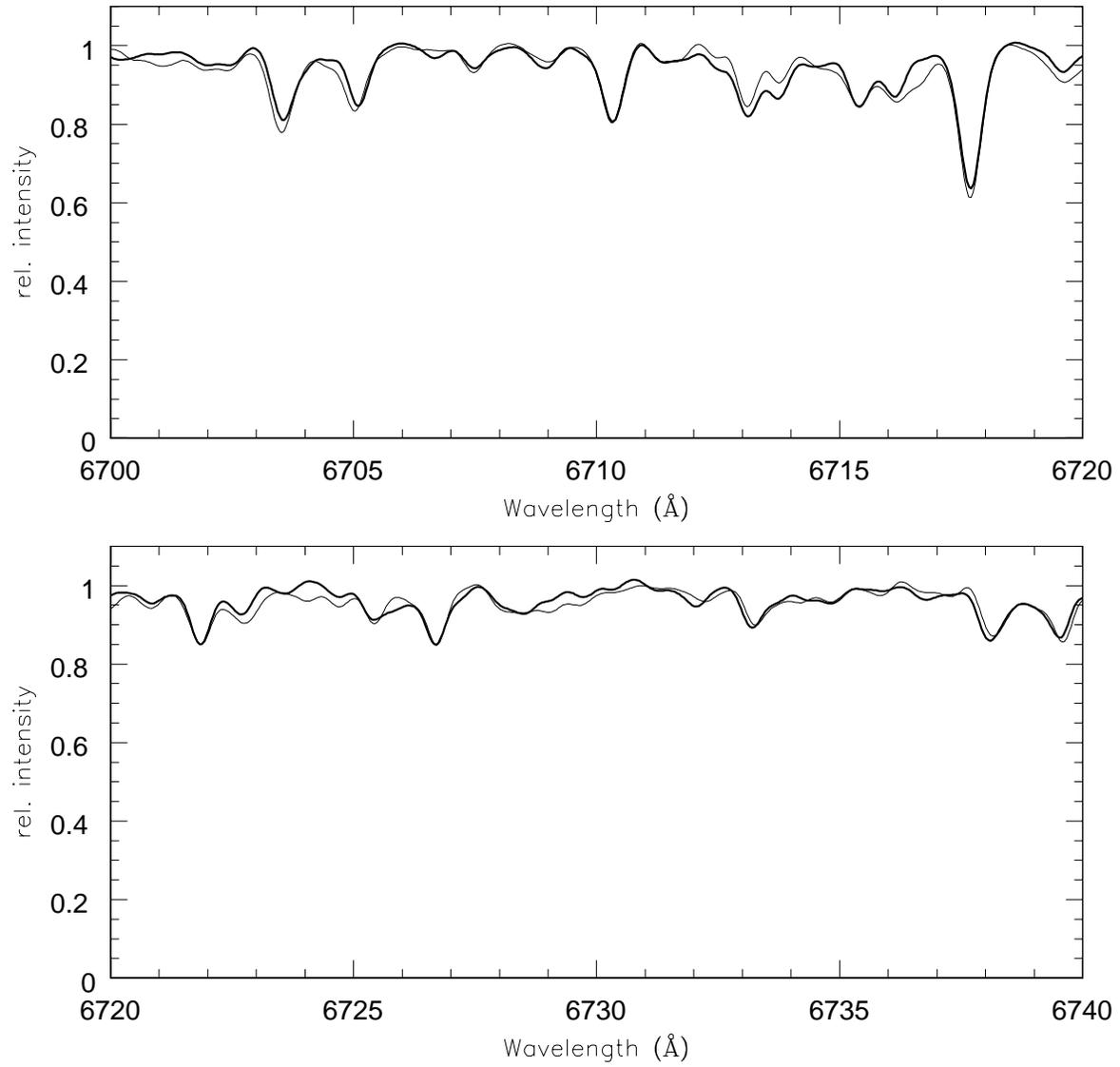}
\caption[carretta.fig7.ps]{Comparison of spectra of two RHB stars in NGC 6553 (star 71, 
see Paper I; light line) and in NGC 6528 (star 3025, from present work; heavy 
line). Spectra have been degraded to the same resolution.
\label{fig7}}
\end{figure}

\begin{figure}
\epsscale{1.0}
\plotone{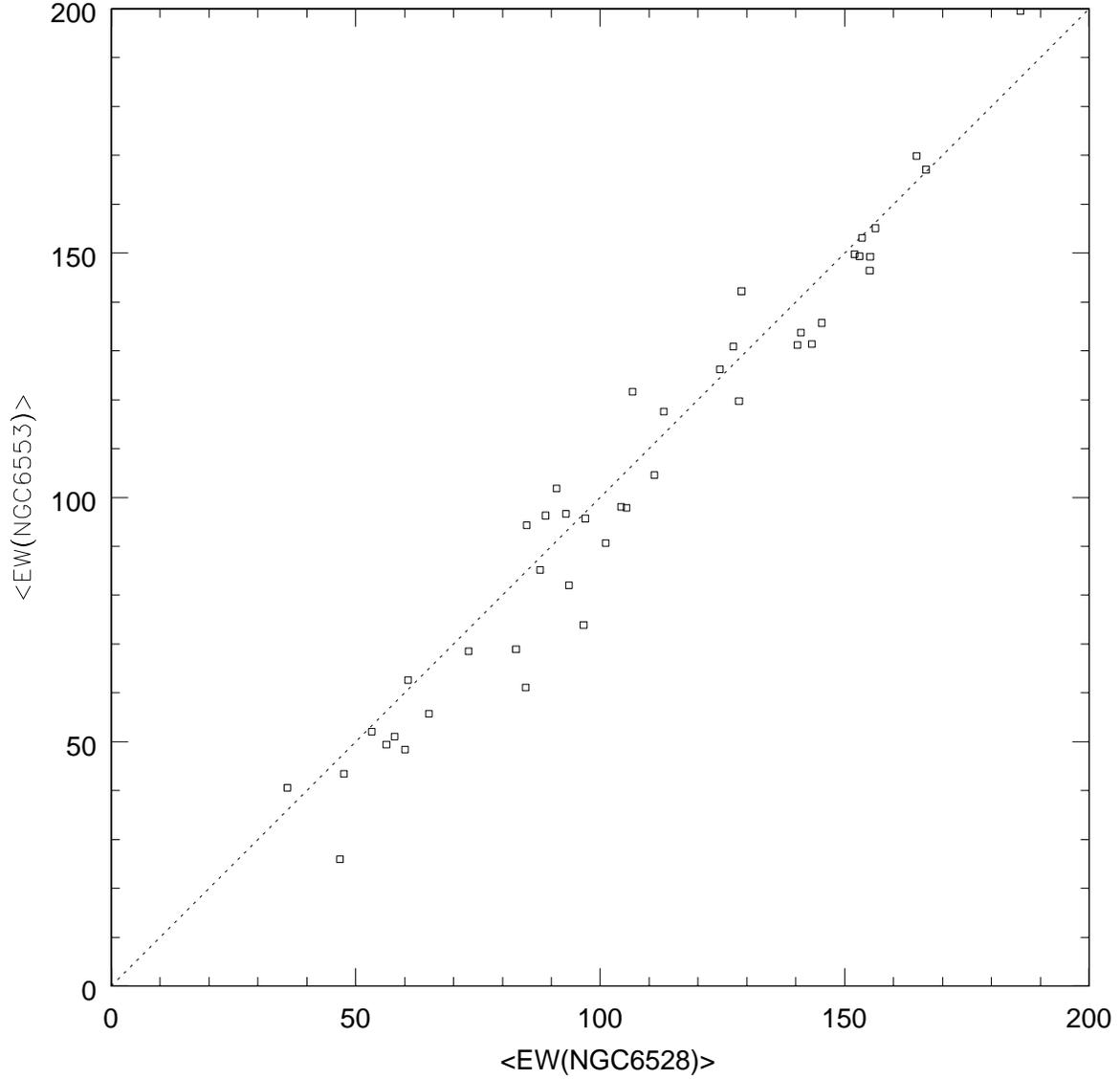}
\caption[carretta.fig8.ps]{Comparison of average EWs for RHB stars in NGC 6553 and NGC 
6528. The average is from the 5 stars studied in Paper I and is shown
when a line is measured in at least 3 stars, for NGC 6553; for NGC 6528, the
average was done using the 3 stars with spectra taken in the 2000 run and only
if a line was measured in at least 2 stars.
\label{fig8}}
\end{figure}

\begin{figure}
\epsscale{1.0}
\plotone{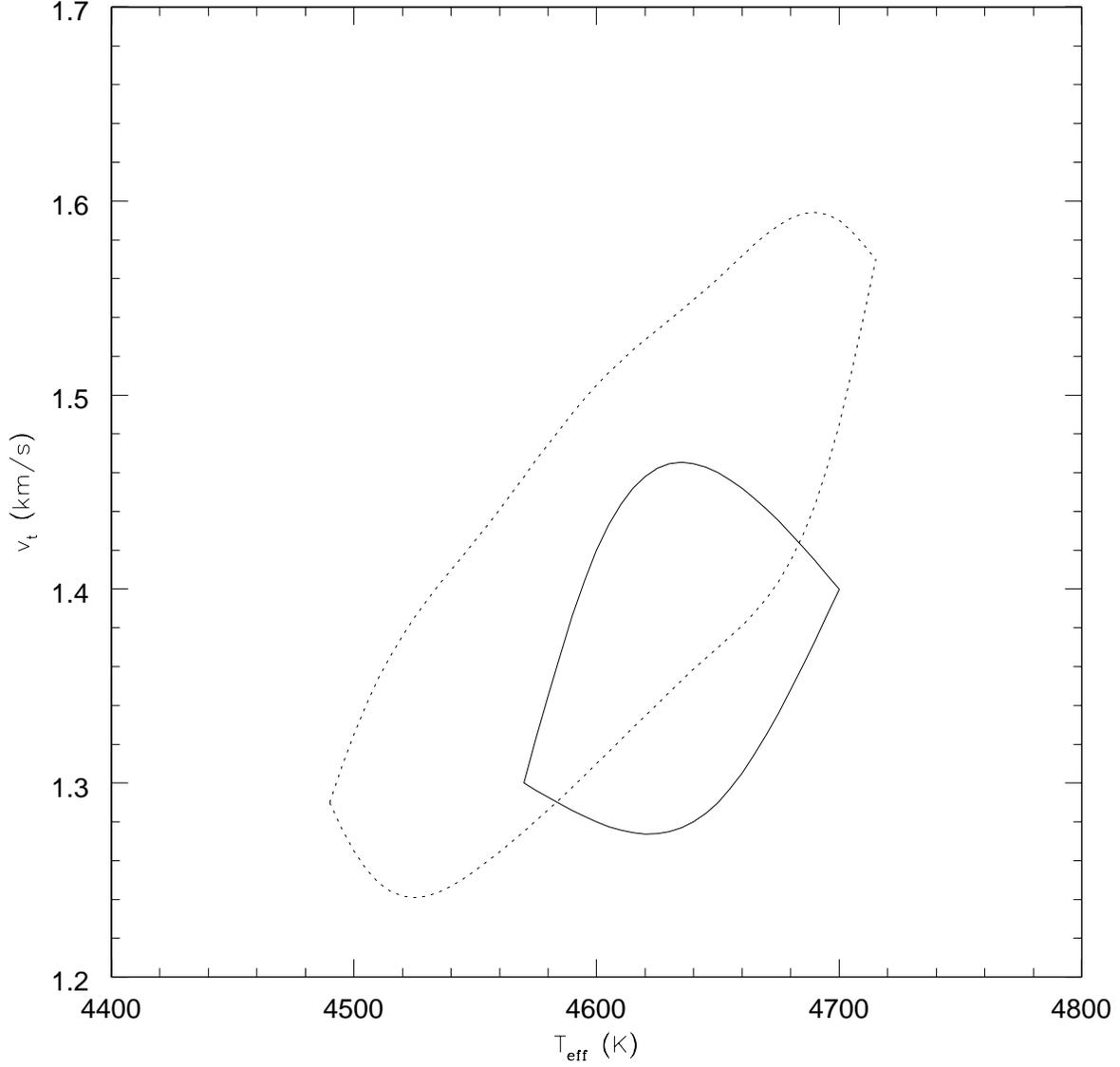}
\caption[carretta.fig9.ps]{Parameter space $v_t$, $T_{\rm eff}$ in the range for RHB
stars. The closed lines define the loci where solutions can be obtained with
errors in slopes of linear regressions of abundances $vs$ \ep\ and of 
expected line strengths within the 1$\sigma$ rms uncertainty.
The dotted line refers to solutions for NGC 6553, while the solid one is for
solutions valid for NGC 6528, for which higher quality spectra and EWs are
available (see text). 
\label{fig9}}
\end{figure}

\begin{figure}
\epsscale{1.0}
\plotone{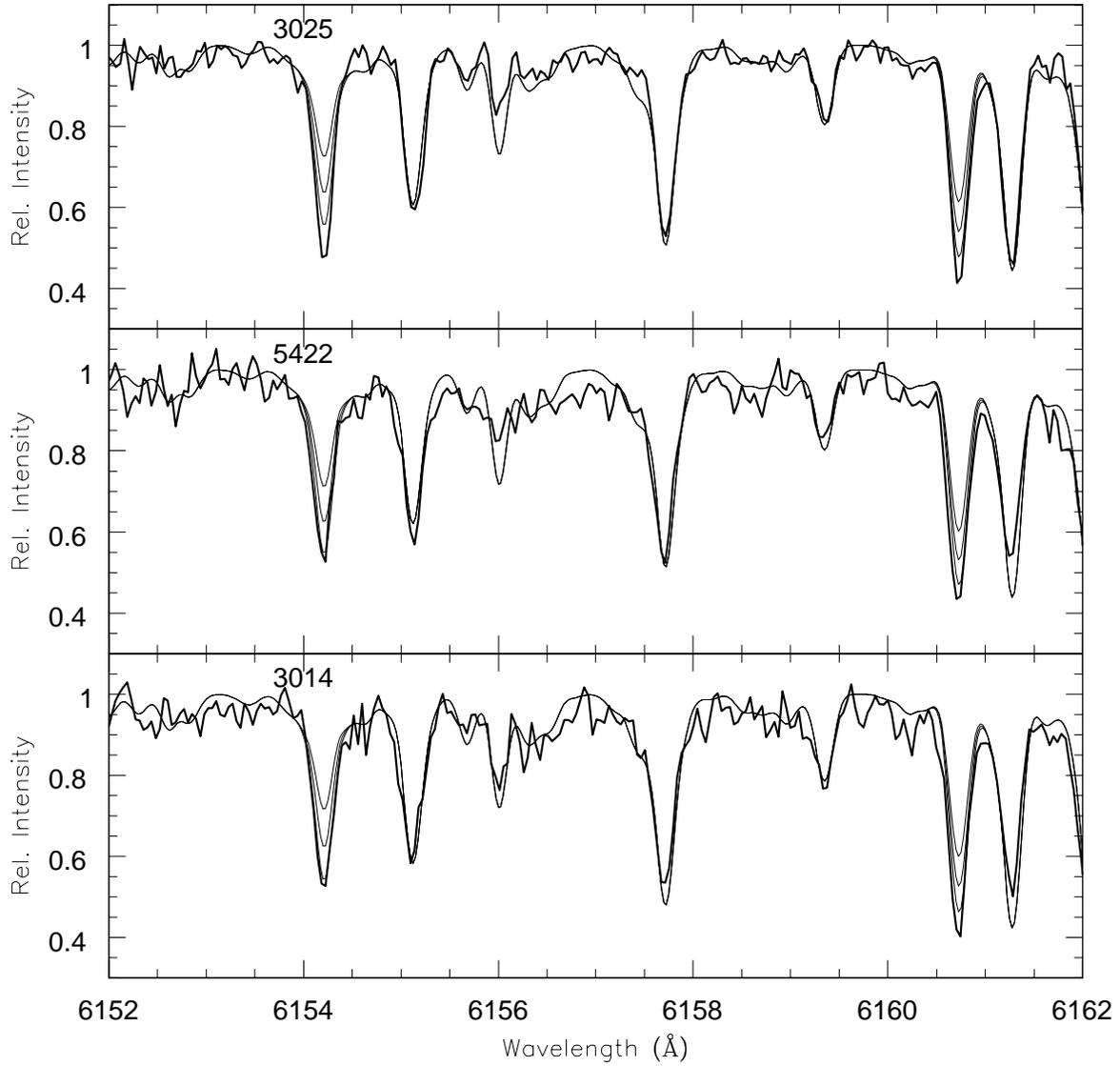}
\caption[carretta.fig10.ps]{Spectrum synthesis of the region of Na doublet at 6154 and 
6160~\AA\ for the RHB stars analyzed in NGC 6528 (only for stars observed in 
2000). Thick lines are the observed spectra, while the thin lines are 
predictions for abundances of [Na/Fe] = 0.0, 0.2 and 0.4 dex.
\label{fig10}}
\end{figure}

\begin{figure}
\epsscale{1.0}
\plotone{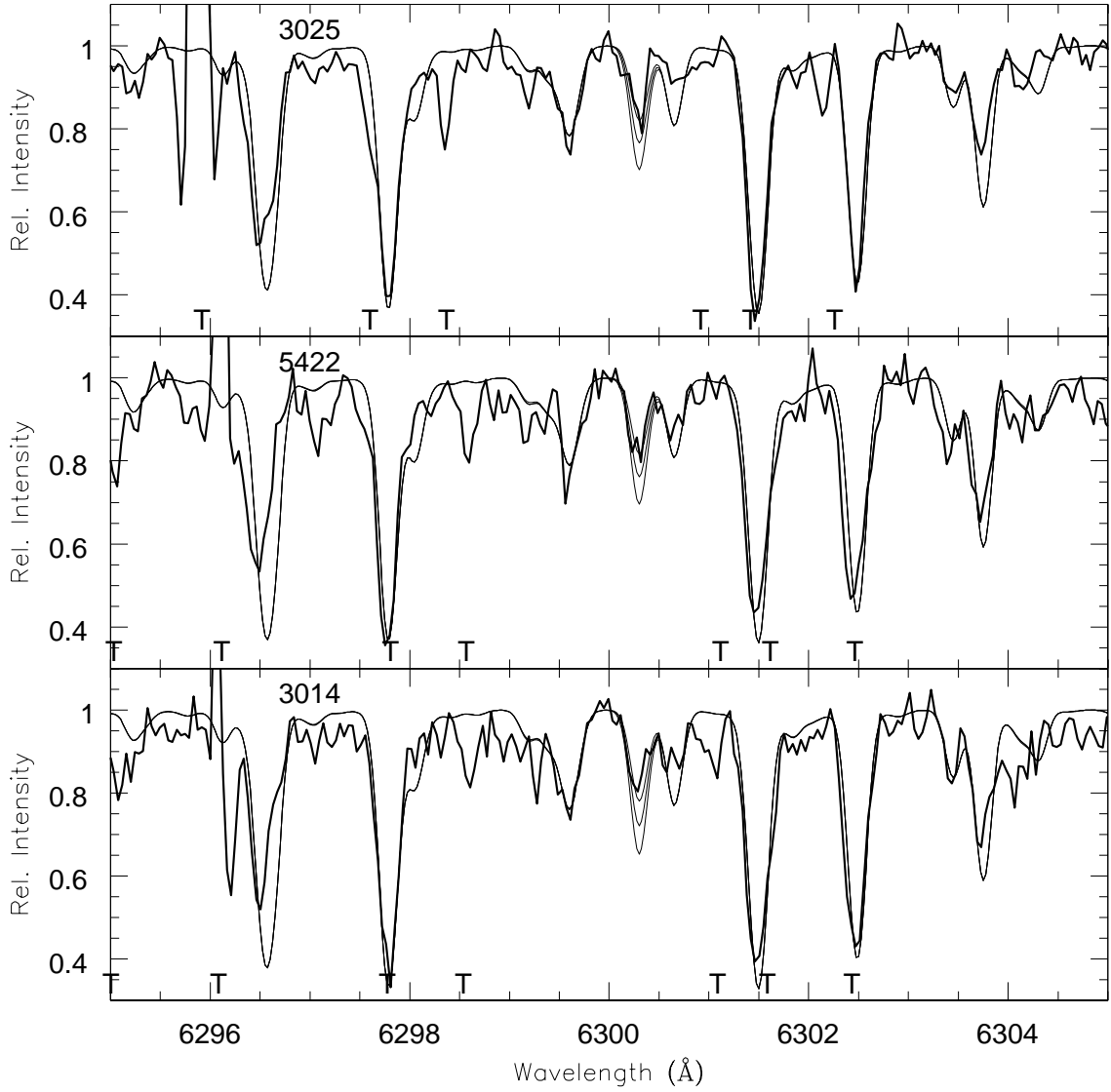}
\caption[carretta.fig11.ps]{Spectrum synthesis of the region of the [O I] forbidden line
at 6300~\AA\ of stars in NGC 6528 observed in the 2000 run. 
Thick lines are the original 
observed spectra, while the thin lines denote predictions for abundances of 
[O/Fe] = $-0.07$, 0.13, and 0.33 dex. The spectra were computed assuming
[C/Fe] = $-0.45$, $-0.30$, and $-0.18$, and [N/Fe]=0; these choices reproduce
the strength of the CN lines. T's mark telluric features.
\label{fig11}}
\end{figure}

\begin{figure}
\epsscale{1.0}
\plotone{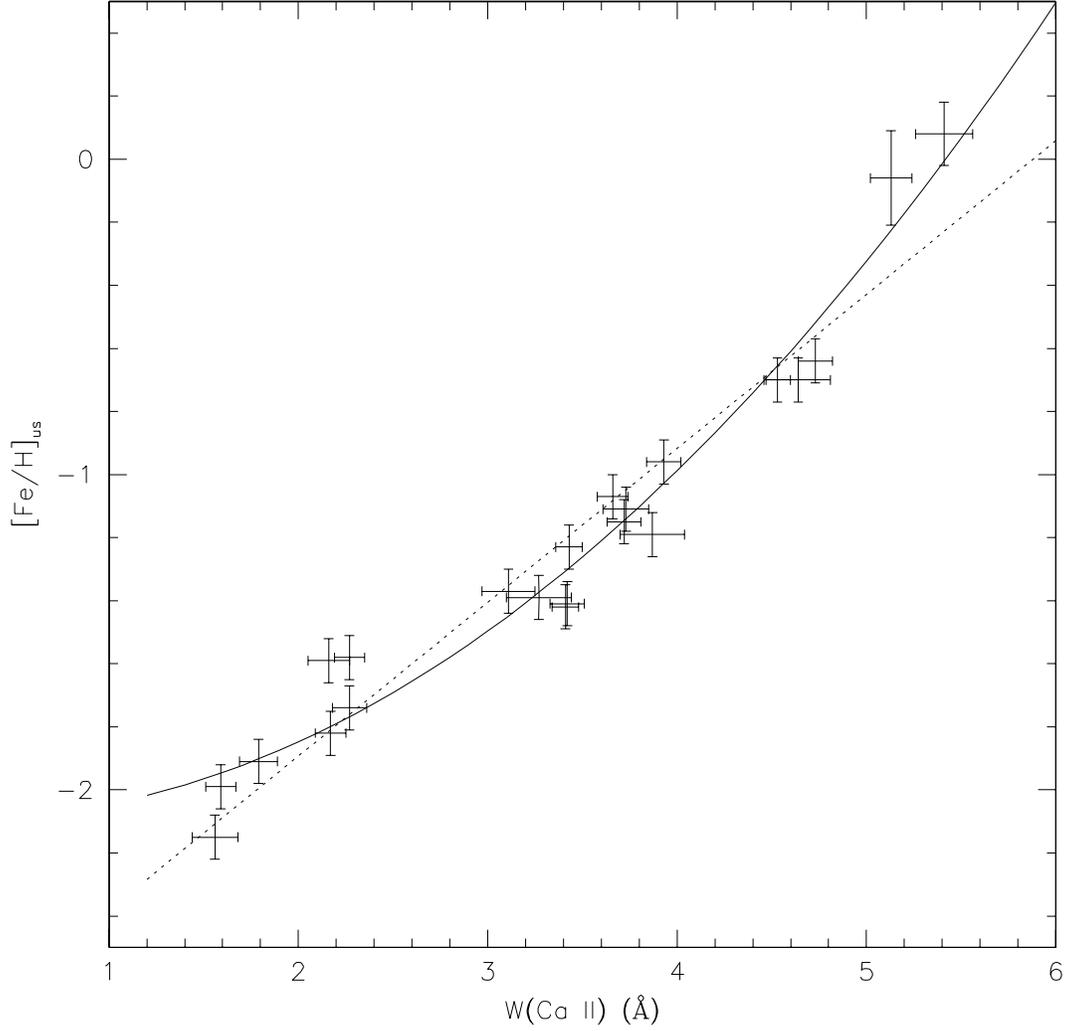}
\caption[carretta.fig12.ps]{The parameter $W(CaII)$ defined by 
Rutledge \etal\ (1997b)
is shown as a function of abundance on the scale of Carretta \& Gratton (1997)
for galactic GCs with high dispersion analyses, as updated by the analysis by
Cohen \etal\ (1999) for NGC 6553 (but see text, Section 4.3) and the present 
study for NGC 6528.
The dashed line is a linear fit, shown for comparison purposes. The solid line
is the transformation adopted (see text).
\label{fig12}}
\end{figure}

\begin{figure}
\epsscale{1.0}
\plotone{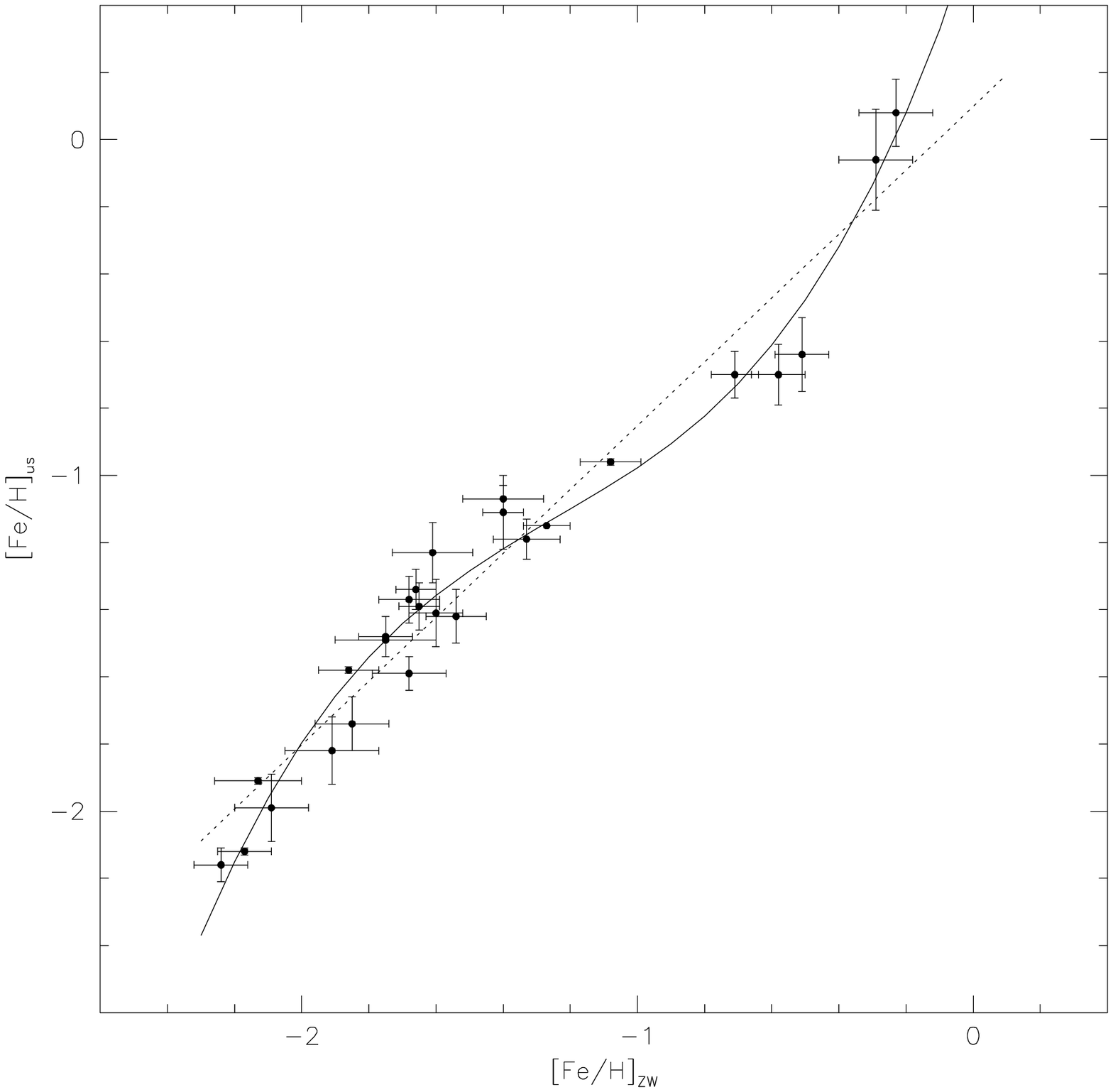}
\caption[carretta.fig13.ps]{The new calibration of the metallicity scale by Zinn \& West 
(1984) against metal abundances from high dispersion spectroscopy. 
The dashed and
solid lines have the same meaning as in the previous figure.
\label{fig13}}
\end{figure}


\clearpage

\begin{references}

Armandroff, T.E. \& Da Costa, G.S., 1991, \aj, 101, 1329

Armandroff, T.E. \&  Zinn, R., 1988, \aj, 96, 92

Barbuy, B., Renzini, A., Ortolani, S., Bica, E. \& Guarnieri, M.~D., 1999,
A\&A, 341, 539

Beaulieu, S.F., Gilmore, G., Elson, R.A.W., Johnson, R.A., Santiago, B., 
Sigurdsson, S. \& Tanvir. N., 2001, \aj, 121, 2618

Bertelli, G., Bressan, A., Chiosi, C., Fagotto, F. \& Nasi, E., 1994, A\&AS, 
106, 275

Cardelli, J.~A., Clayton, G.~C. \& Mathis, J.~S., 1989, \apj, 345, 245

Carretta, E. \& Gratton, R.G., 1997, A\&A Supl, 121, 95

Cohen, J.~G., 1983, \apj, 270, 654 

Cohen, J.~G., Gratton, R.G., Behr, B.B. \& Carretta, E. 1999, ApJ, 523, 739 
(Paper I)

Cohen, J.~G. \& Sleeper, E.~C., 1995, \aj, 109, 242

Cook, K., Mateo, M., Olszewski, E.W., Vost, S.S., Stubbs, C. \& Diercks, A.,
1999, \pasp, 111, 306

C\^ot\'e, P., 1999, \aj, 118, 406

C\^ot\'e, P., Mateo, M., Olszewski, E.W. \& Cook, K.H., 1999, \apj, 526, 147

Davidge, T.J., 1999, astro-ph/9909408

Davis, M. \& Tonry, J.L., 1977, \aj, 84, 1511

Feast, M.W., Robertson, B.S.C., \& Black, C., 1980, MNRAS, 190, 227

Feltzig, S. \& Gilmore, G. 2000, Astro-ph/0002123

Gratton, R.~G., 1987, MNRAS, 224, 175

Gratton, R.~G., 1989, A\& A, 208, 171

Gratton, R.~G., Carretta, E. \& Castelli, F., 1996, A\& A, 314, 191

Gratton, R.~G., Carretta, E., Eriksson, K. \& Gustafsson, B., 1999, A\&A, 
350, 955

Gratton, R.G. et al., 2001, A\& A, 369, 87

Gratton, R.~G., Sneden, C., 1991, A\& A, 241, 501

Harris, W.E. 1996, AJ, 112, 1487

Ivans, I.I., Sneden, C., Kraft, R.P., Suntzeff, N.B., Smith, V.V., Langer, G.E.
  \& Fullbright, J.P. 1999, \aj, 118, 1273

Kraft, R.~P.,  Sneden, C., G Smith, G.~H., Shetrone, M.~D. \& 
Fulbright, J., 1998, \aj, 115, 1500

Kurucz, R.~L., 1992, CD-ROM 13

Magain, P., 1984, A\&A, 134, 189

Mateo, M., Olszewski, E.W., Vogt, S.S. \& Keane, M.J., 1998, \aj,
116, 2315

McCarthy, J.~K., 1988, PhD thesis, California Institute of Technology

McWilliam, A. \& Rich, R.~M., 1994, \apjs, 91, 749

Minniti, D., 1995, A\&A Suppl., 113, 299

Ortolani, S., 1999, invited talk, in The Chemical Evolution of the Galaxy: 
Stars versus Clusters, Vulcano, September 1999, in press

Ortolani, S., Barbuy, B. \& Bica, E., 1990, A\&A, 236, 362   

Ortolani, S., Bica, E. \& Barbuy, B., 1992, A\&A Suppl, 92, 441

Ortolani, S., Renzini, A., Gilmozzi, R., Marconi, G., Barbuy, B., Bica, E. \&
 Rich, R.M. 1995, Nature, 377, 701

Reed, B.C., Hesser, J.~E. \& Shawl, S.~J., \pasp, 100, 545

Rich, R.~M., 1988, \aj, 95, 828

Rich, R.~M. \& McWilliam, A., 2000, to appear in SPIE, 4005 (ed. J.Bergeron).
(Astro-ph/0005113)

Richtler, T., Grebel, E.K., Subramaniam, A., Sagar, R., 1998, A\&AS, 127, 169

Rutledge, G.~A., Hesser, J.~E., Stetson, P.~B., Mateo, M., Simard, L., 
Bolte, M., Friel, E.~D. \& Copin, Y., 1997a, \pasp, 109, 883 

Rutledge, G.~A., Hesser, J.~E.\& Stetson, P.~B., 1997b, \pasp, 109, 907

Sadler, E.~M., Rich, R.~M., \& Terndrup, D.~M., 1996, \aj, 112, 171

Salasnich, B., Girardi, L., Weiss, A. \& Chiosi, C., 2000, in preparation

Sarajedini, A. 1994, \aj, 107, 618

Sharples, R., Walker, A., \& Cropper, M., 1990, MNRAS, 246, 54

Shortridge, K. 1988, ``The Figaro Manual Version 2.4''

Smith, G., \& Ragget, D.~St.~J., 1981, J. Phys. B. At. Mol. Phys., 14, 4015

Stanek, K.,1996, \apjl, 460, L37

Terndrup, D.~M., Sadler, E.~M., Rich, R.~M., 1995, \aj, 110, 1774

Ventura, P., D'Antona, F., Gratton, R.G.,2001, \apjl, 550, L65

Vogt, S.E., Allen, S., Bigelow, B., Bresee, L., Brown, B., Cantrall, T.,
Conrad, A., Couture, M., Delaney, C., Epps, H., Hilyard, D., Hilyard, D.,
Horn, E., Jern, N., Kanto, D., Keane, M., Kibrick, R., Lewis, J.,
Osborne, C., Osborne, J., Pardeilhan, G., Pfister, T., Ricketts, T.,
Robinson, L., Stover, R., Tucker, D., Ward, J. \& Wei, M.,
1994, SPIE, 2198, 362

Zinn, R.~J., 1980, \apjs, 42, 19

Zinn, R. \& West, M. 1984, \apjs, 55, 45

\end{references}
\end{document}